\documentclass[a4paper,11pt]{article}
\usepackage[utf8x]{inputenc}
\usepackage{amssymb,amsfonts,amstext,amsmath,graphicx,epic,amsthm,color,times}

\title{{ Parametrics Resonances of a Forced Modified Rayleigh-Duffing Oscillator}
}
\author{C. H. Miwadinou\footnote{clement.miwadinou@imsp-uac.org,hodevewan@yahoo.fr}, A. V. Monwanou\footnote{movins2008@yahoo.fr} and J. B. Chabi Orou\footnote{Author to whom correspondence should be addressed: jchabi@yahoo.fr}}

\begin{document}

\maketitle Institut de Math\'ematiques et de Sciences Physiques, BP: 613 Porto Novo, B\'enin

\begin{abstract}
We investigate in this paper the superharmonic and subharmonic resonances of forced modified Rayleigh-Duffing oscillator. We analyse this equation by method
of multiple scales and we obtain superharmonic, subharmonic resonances order-two and order-three and primary resonance. We obtain also regions where steady-state subharmonic responses
exist. We also use the amplitude-frequency curve for demonstrate the effect of various parameters on the response of the system.
Finally, we focus our attention on chaotic motion of this oscillator by simulation and we obtain that this 
oscillator is chaotic for certains values for natural and excitation frequency but chaotic motion it is not the
same in subharmonic and superharmonic cases.

\end{abstract}

{\bf keywords}: Parametric resonance, superharmonic, subharmonic, forced modified Rayleigh-Duffing oscillator, multiple scales method,
   chaotic behavior.

\section{Introduction}
Many problems in physics, chemistry, biology, etc., are related to nonlinear self-excited oscillators \cite{2}. For example, the self-excited oscillations
 in bridges and airplane wings, the beating of a heart, and the nonlinear model of a machine tool chatter \cite{10}. A self-excited oscillator is a system 
which has some external source of energy upon which it can be drawn. Self-excited systems have a long history in the field of mechanics \cite{11,12}.
 One of key problems in the theory of nonlinear oscillations is a search of possibilities to estimate their amplitude and period analytically. 
Parametric perturbations are characterised by parameters periodically in time changing and they are described by homogeneous differential 
equations of motion. Many works on self-excited, parametrically and externally excited are well known and deeply investigated in the 
literature separately. Minorski \cite{14} is one of the first authors considering the interaction between two different types of perturbations. 
Warminski \cite{16} emphasizes the differences in modelling ideal and non-ideal systems for a chosen class of self-excited, parametric and externally
 excited vibrations. Many of those studies lead to the parametric excitation combined with self-excited system and subjected to an external force
which quite often take the form 
\begin{equation}
 \ddot{x}+\eta(x;\dot{x})\dot{x}+(1-\mu\cos2\omega t)(x+\alpha x^3)=F\cos\omega t,\label{eq.y}
\end{equation}
where $\eta(x;\dot{x})$ is a nonlinear damping function. The effect of nonlinear damping on a nonlinear oscillator was investigated
previously in \cite{15}, showing among other things how it affected the evolution of fractalisation of phase space.

Autoparametric resonance plays an important part in nonlinear engineering while posing interesting mathematical challenges. The linear dynamics 
is already nontrivial whereas the nonlinear dynamics of such systems is extremely rich and largely unexplored \cite{5}. 
Tina Marie Morrison in his thesis \cite{4}, have investigated the dynamics of a system consisting of a simple harmonic oscillator with small nonlinearity,
 damping  and parametric forcing in the neighborhood of 2:1 resonance  near a Hopf bifurcation:
\begin{equation}
 \ddot{z}+\epsilon A\dot{z}+(1+\epsilon k_1+\epsilon B\cos2t)z+\epsilon(\beta_1z^3+\beta_2z^2\dot{z}+\beta_3z\dot{z}^2+\beta_4\dot{z}^3)=0
\end{equation}
 Venkatanarayanan Ramakrishnan and Brian F Feeny are particulary study in \cite{1} the resonances of the forced nonlinear Mathieu equation.

In the present work we consider the modified Rayleigh-Duffing oscillator modelled by following equation:
 \begin{eqnarray}
 &&\ddot{x} + \epsilon\mu (1-\dot{x}^2)\dot{x}+\epsilon\beta\dot{x}^2+\epsilon k_1\dot{x}x+\epsilon k_2\dot{x}^2 x +(\omega^2+\epsilon \alpha\cos\Omega t)x\cr
&&+\epsilon\lambda x^3=F\cos\Omega t. \label{eq.0}
\end{eqnarray}
Our interest in understanding the behavior of this equation  is motivated by two applications. The first is a model of the El Ni$\tilde{n}$o Southern 
Oscillation (ENSO) coupled tropical ocean-atmosphere weather phenomenon \cite{39,40}  in which the state variables are temperature and depth of 
a region of the ocean called the thermocline. The annual seasonal cycle is the parametric excitation. The model exhibits a Hopf bifurcation in the
 absence of parametric excitation.
The second application involves a MEMS device \cite{41,42} consisting of a $30 \mu m$ diameter silicon disk which can be made to vibrate by heating it 
with a laser beam resulting in a Hopf bifurcation. The parametric excitation is provided by making the laser beam intensity vary periodically in time.

We focus our attention on the study of the differents resonances which can exist in the forced parametric modified Rayleigh-Duffiing oscillator. 
We seek approximate solutions to equation (\ref{eq.0})  by using the method of multiple scales (MMS) and we find the peak amplitude of resonances 
phenomenon. We study the effects of certains parameters of this oscillator on these differents resonances. Finally, we study
the chaotic motion of this oscillator by simulation in the subharmonic and superharmonic regions.

\section{Resonances of the forced modified Rayleigh-Duffing oscillator}
We use the method of multiple scales (MMS) to seek approximate solutions to equation (\ref{eq.0}). The analysis reveals 
the existence of various superharmonic and subharmonic resonances. The method of multiple scales supposed that the approximate
 steady solution of first order for eq.(\ref{eq.0}) in the form \cite{35}
\begin{eqnarray}
 x(t,\epsilon)=x_0(T_0,T_1)+\epsilon x_1(T_0,T_1)+........,\label{eq.1}
\end{eqnarray}
where  $T_0=t, T_n=\epsilon^nT_0 $. Then $\frac{d}{dt}=D_0+\epsilon D_1, \frac{d^2}{dt^2}=D_0^2+2\epsilon D_0D_1+... $ with 
$D_n^m=\frac{\partial^m}{\partial T_n^m}.$

Substituting eq.(\ref{eq.1}) into eq.(\ref{eq.0}) and equating the coefficients of the same power of small parameter $ \epsilon$, one 
obtains\\
In order $\epsilon^o$, 
\begin{eqnarray}
 D_0^2x_0+\omega^2x_0=F\cos\Omega T_0 .\label{eq.2}
\end{eqnarray}
In order $\epsilon^1$,
\begin{eqnarray}
 D_0^2x_1+\omega^2x_1&=&-2D_0D_1x_0-\mu D_0x_0+\mu(D_0x_0)^3-\beta(D_0x_0)^2-\cr
&&\alpha x_0\cos\Omega T_0-k_1x_0D_0x_0-k_2x_0(D_0x_0)^2-\lambda x_0^3 .\label{eq.3}
\end{eqnarray}

The solution for eq.(\ref{eq.2}) is 
\begin{eqnarray}
 x_0=Ae^{i\omega T_0}+\Lambda e^{i\Omega T_0}+cc,   \label{eq.4} 
\end{eqnarray}
where 
\begin{eqnarray}
\Lambda=\frac{F}{2(\omega^2-\Omega^2)}, A=\frac{1}{2}ae^{i\theta}, \label{eq.x}
\end{eqnarray}
cc stands for complex conjugate of preceding terms.
Substituting the solution $x_0$ from eq.(\ref{eq.4}) into eq.(\ref{eq.3}), we are expanded the terms on the right hand side. We obtain 
\begin{eqnarray}
 D_0^2x_1+\omega^2x_1&=& [-2i\omega A'-i\mu\omega A+3i\mu\omega^3A^2\bar{A}+6i\mu\omega\Omega^2 A\Lambda^2-3\lambda A^2\bar{A}-\cr
&&6\lambda A\Lambda^2-k_2\omega^2A^2\bar{A}-2k_2\Omega^2A\Lambda^2]e^{i\omega T_0}+[-i\mu \Lambda\Omega+\cr
&&6i\mu\omega^2\Omega A\bar{A}\Lambda+3i\Omega^3\Lambda^3-6\lambda\Lambda A\bar{A}-3\lambda\Lambda^3-2k_2\omega^2A\bar{A}\Lambda-\cr
&&k_2\Lambda^3\Omega^3]e^{i\Omega T_0}+[-i\mu\omega^3A^3-\lambda A^3+k_2\omega^2A^3]e^{i3\omega T_0}+\cr
&&[-i\mu\Lambda^3\omega^3-\lambda\Lambda^3+k_2\Omega^2\Lambda^3]e^{i3\Omega T_0}+[-3i\mu\omega^2\Omega\bar{A}^2\Lambda-\cr
&&3\lambda\bar{A}^2\Lambda+k_2\omega^2\bar{A}^2\Lambda-2k_2\omega\Omega\bar{A}^2\Lambda]e^{i(-2\omega+\Omega )T_0}+\cr
&&[-\frac{1}{2}\alpha\bar{A}-ik_1\Omega\Lambda\bar{A}+ik_1\omega\Lambda\bar{A}-2\beta\omega\Omega\bar{A}\Lambda]e^{i(-\omega+\Omega )T_0}+\cr
&&[-\frac{1}{2}\alpha\Lambda-ik_1\Omega\Lambda^2+\beta\Lambda^2\Omega^2]e^{i2\Omega T_0}+\beta A^2e^{i2\omega T_0}-\cr
&&2\beta\omega^2A\bar{A}-2\beta\Lambda^2\Omega^2+cc+NST,\label{eq.5}
\end{eqnarray}
where NST is non resonance terms.
We need to eliminate coefficients of $e^{i\omega T_0} $ that constitute the secular terms and would make the solutions unbounded. The solvability 
condition is thus set by equating the coefficients of $e^{i\omega T_0}$ terms to zero.

\subsection{Superharmonic resonances}
 In this case, we consider first $2\Omega=\omega+\epsilon\sigma$ and after $3\Omega=\omega+\epsilon\sigma$ where $\sigma $ is a detuning parameter.
\subsubsection{$2\Omega=\omega+\epsilon\sigma$}
If $2\Omega=\omega+\epsilon\sigma$, the condition for elimination of secular terms in eq.(\ref{eq.5}) is 
\begin{eqnarray}
&&-2i\omega A'-i\mu\omega A+3i\mu\omega^3A^2\bar{A}+6i\mu\omega\Omega^2 A\Lambda^2-3\lambda A^2\bar{A}-6\lambda A\Lambda^2-\cr
&&k_2\omega^2A^2\bar{A}-2k_2\Omega^2A\Lambda^2+(-\frac{1}{2}\alpha\Lambda-ik_1\Omega\Lambda^2+\beta\Lambda^2\Omega^2)e^{i\sigma T_1}=0, \label{eq.6}
\end{eqnarray}
with $T_1=\epsilon T_0 $. To this order, $A$ is considered to be a function to $ T_1$ only. Then, substituting the polar form eq.(\ref{eq.x}) into 
 eq.(\ref{eq.6}) and equating the real and imaginary parts, one gets
\begin{eqnarray}
 a'&=&(-\frac{1}{2}+3\Omega^2\Lambda^2)\mu a+\frac{3}{8}\mu\omega^2 a^3-\frac{\Omega k_1\Lambda^2}{\omega}\cos(\sigma T_1-\theta)+ \cr
&&\frac{(-\frac{1}{2}\alpha\Lambda+\beta\Omega^2\Lambda^2)}{\omega}\sin(\sigma T_1-\theta), \label{eq.7}
\end{eqnarray}
\begin{eqnarray}
 a\theta'&=&(3\lambda+k_2\Omega^2)\frac{\Lambda^2}{\omega}a+\frac{1}{8}(3\lambda+k_2\omega^2)\frac{a^3}{\omega}-\frac{\Omega k_1\Lambda^2}{\omega}\sin(\sigma T_1-\theta)-\cr
&&\frac{(-\frac{1}{2}\alpha\Lambda+\beta\Omega^2\Lambda^2)}{\omega}\cos(\sigma T_1-\theta). \label{eq.8}
\end{eqnarray}
Letting $\gamma=\sigma T_1-\theta $, eq.(\ref{eq.7}) and eq.(\ref{eq.8}) can be written to 

\begin{eqnarray}
 a'&=&(-\frac{1}{2}+3\Omega^2\Lambda^2)\mu a+\frac{3}{8}\mu\omega^2 a^3-\frac{\Omega k_1\Lambda^2}{\omega}\cos\gamma+ \cr
&&\frac{(-\frac{1}{2}\alpha\Lambda+\beta\Omega^2\Lambda^2)}{\omega}\sin\gamma, \label{eq.9}
\end{eqnarray}
\begin{eqnarray}
 a\gamma'&=& a\sigma-(3\lambda+k_2\Omega^2)\frac{\Lambda^2}{\omega}a-\frac{1}{8}(3\lambda+k_2\omega^2)\frac{a^3}{\omega}+\frac{\Omega k_1\Lambda^2}{\omega}\sin\gamma+\cr
&&\frac{(-\frac{1}{2}\alpha\Lambda+\beta\Omega^2\Lambda^2)}{\omega}\cos\gamma. \label{eq.10}
\end{eqnarray}
Putting $a'=\theta'=0$ to find the stable period solution. We obtain
 \begin{eqnarray}
 &&(-\frac{1}{2}+3\Omega^2\Lambda^2 +\frac{3}{8}\omega^2 a^2)\mu a=\frac{\Omega k_1\Lambda^2}{\omega}\cos\gamma- \cr
&&              \frac{(-\frac{1}{2}\alpha\Lambda+\beta\Omega^2\Lambda^2)}{\omega}\sin\gamma, \label{eq.11}
\end{eqnarray}
\begin{eqnarray}
&&[\sigma-(3\lambda+k_2\Omega^2)\frac{\Lambda^2}{\omega}-\frac{1}{8}(3\lambda+k_2\omega^2)\frac{a^2}{\omega}]a=-\frac{\Omega k_1\Lambda^2}{\omega}\sin\gamma-\cr
&&\frac{(-\frac{1}{2}\alpha\Lambda+\beta\Omega^2\Lambda^2)}{\omega}\cos\gamma. \label{eq.12}
\end{eqnarray}
Considering these equations eq.(\ref{eq.11}) and eq.(\ref{eq.12}), the frequency-response curve for superharmonic resonance is 
 \begin{eqnarray}
&&[[\sigma-(3\lambda+k_2\Omega^2)\frac{\Lambda^2}{\omega}-\frac{1}{8}(3\lambda+k_2\omega^2)\frac{a^2}{\omega}]^2+\cr
&&(-\frac{1}{2}+3\Omega^2\Lambda^2 +\frac{3}{8}\omega^2 a^2)^2\mu^2]a^2=\frac{\Omega^2 k_1^2\Lambda^4}{\omega^2}+\frac{(-\frac{1}{2}\alpha\Lambda+
\beta\Omega^2\Lambda^2)^2}{\omega^2}. \label{eq.13}
 \end{eqnarray}

At steady-state the relationship between the response amplitude and the detuning parameter $ \sigma$ is 
\begin{eqnarray}
 \sigma= (3\lambda+k_2\Omega^2)\frac{\Lambda^2}{\omega}+\frac{1}{8}(3\lambda+k_2\omega^2)\frac{a^2}{\omega}\pm\cr
[\frac{\Omega^2 k_1^2\Lambda^4}{\omega^2a^2}+\frac{(-\frac{1}{2}\alpha\Lambda+\beta\Omega^2\Lambda^2)^2}{\omega^2a^2}-(-\frac{1}{2}+3\Omega^2\Lambda^2 +
\frac{3}{8}\omega^2 a^2)^2\mu^2]^{\frac{1}{2}}. \label{eq.14}
\end{eqnarray}
The peak amplitude would be  verify the following equation:
\begin{eqnarray}
 \frac{\Omega^2 k_1^2\Lambda^4}{\omega^2a_p^2}+\frac{(-\frac{1}{2}\alpha\Lambda+\beta\Omega^2\Lambda^2)^2}{\omega^2a_p^2}=(-\frac{1}{2}+3\Omega^2\Lambda^2 +
\frac{3}{8}\omega^2 a_p^2)^2\mu^2. \label{eq.15}
\end{eqnarray}
We obtain that the corresponding value of $ \sigma$ is 
\begin{eqnarray}
 \sigma_p=(3\lambda+k_2\Omega^2)\frac{\Lambda^2}{\omega}+\frac{1}{8}(3\lambda+k_2\omega^2)\frac{a_p^2}{\omega}. \label{eq.16}
\end{eqnarray}
We can conclude  the following:\\
$\star$ the peak value is independent of $k_2$ and $\lambda $,\\
$\star$ parameters of modified Rayleigh-Duffing oscillator affect the peak location and as they increase, $|\sigma_p| $ increases.
Now we plot the  frequency-response curve from eq.(\ref{eq.13}). In Fig. 1, the frequency-response curve are plotted for fixed values 
of linear and 
nonlinear parameters. This curve shows that the amplitude of the resonance frequency when augment the external exciting force to which the 
order-two resonance superharmonic increases. We also note that the peak of the resonance curve becomes less sharp as the frequency increases.  
\begin{figure}[htbp]
\begin{center}
 \includegraphics[width=12cm, height=6cm]{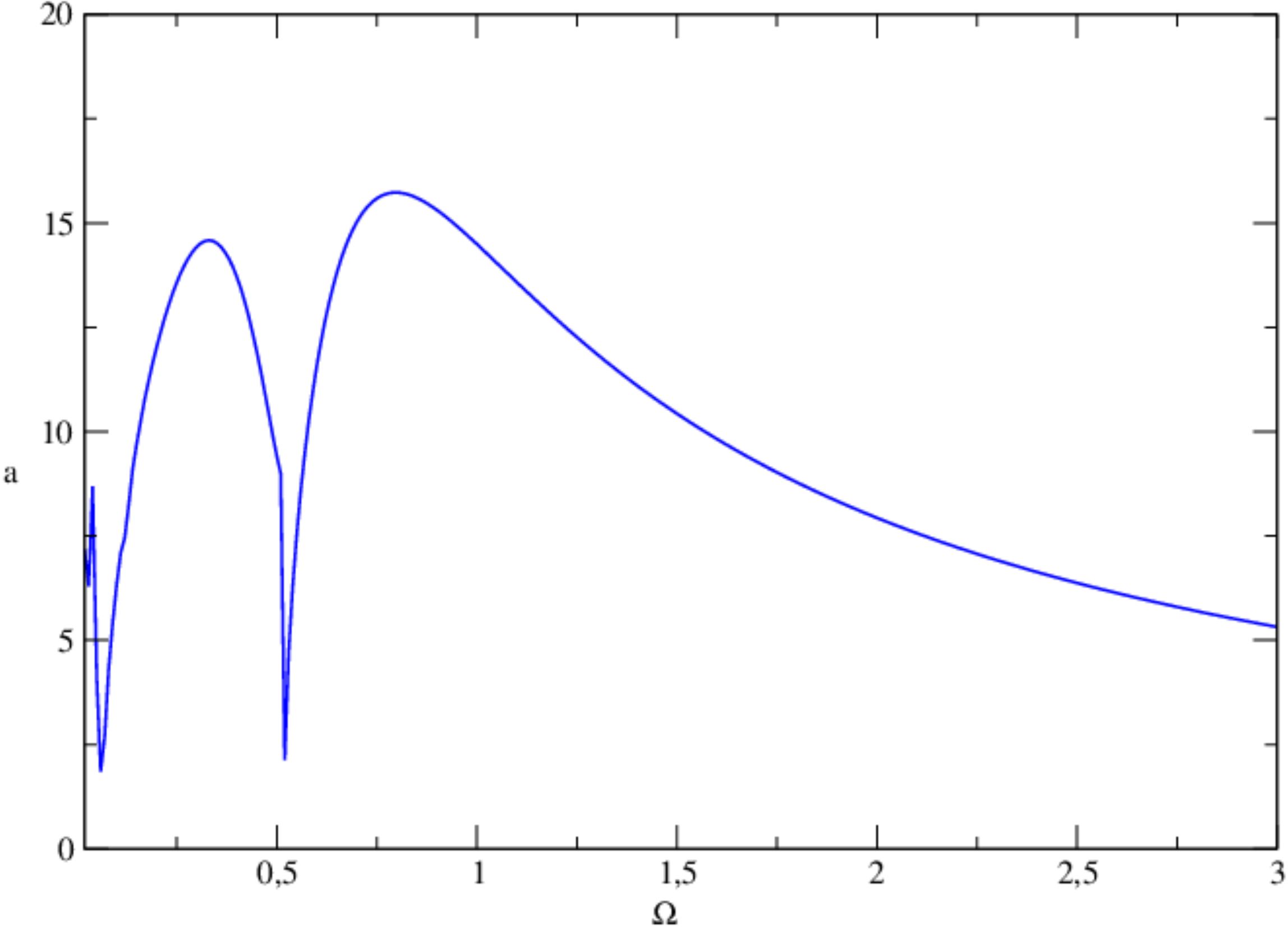}
\end{center}
\caption{superharmonic resonance in the space$(a,\Omega )$ for $\mu=0.5; \alpha=0.5; \lambda=0.5; \beta=0.1; k_2=0.5; k_1=0.1;F=0.01; \epsilon=0.005 $.}
\end{figure}
\newpage
\subsubsection{$3\Omega=\omega+\epsilon\sigma$}
If $3\Omega=\omega+\epsilon\sigma$, the first term and the term which have $ 3\Omega T_0$ as an exponential argument of the right member of
 eq.(\ref{eq.5}) are the secular terms. The condition for the elimination of secular terms is
\begin{eqnarray}
&& -2i\omega A'-i\mu\omega A+3i\mu\omega^3A^2\bar{A}+6i\mu\omega\Omega^2 A\Lambda^2-3\lambda A^2\bar{A}-6\lambda A\Lambda^2-\cr
&&k_2\omega^2A^2\bar{A}-2k_2\Omega^2A\Lambda^2+(-i\mu\Lambda^3\omega^3-\lambda\Lambda^3+k_2\Omega^2\Lambda^3)e^{i\sigma T_1}=0, \label{eq.17}
\end{eqnarray}

Following the analysis done in the previous section for the $2\Omega$ superharmonic resonance, we substitute for $ A$ and 
separate the equation into real and imaginary parts. Using $\gamma=\sigma T_1-\theta $, we arrive at a homogenous set of equations in $\gamma$ and $a$ 
\begin{eqnarray}
 a'=(-\frac{1}{2}+3\Omega^2\Lambda^2)\mu a+\frac{3}{8}\mu\omega^2a^3-\mu\omega^2\Lambda^3\cos\gamma-
\frac{(\lambda-k_2\Omega^2)}{\omega}\Lambda^3\sin\gamma, \label{eq.18}
\end{eqnarray}
and
\begin{eqnarray}
 a\gamma'&=& a\sigma-(3\lambda+k_2\Omega^2)\frac{\Lambda^2}{\omega}a-\frac{1}{8}(3\lambda+k_2\omega^2)\frac{a^3}{\omega}+\mu\omega^2\Lambda^3\sin\gamma-\cr
&&\frac{(\lambda-k_2\Omega^2)}{\omega}\Lambda^3\sin\gamma. \label{eq.19}
\end{eqnarray}

For steady-state solutions $a'=\gamma'=0$, which is satisfied if 
\begin{eqnarray}
 &&[[\sigma-(3\lambda+k_2\Omega^2)\frac{\Lambda^2}{\omega}-\frac{1}{8}(3\lambda+k_2\omega^2)\frac{a^2}{\omega}]^2+\cr
&&(-\frac{1}{2}+3\Omega^2\Lambda^2+\frac{3}{8}\omega^2a^2)^2\mu^2]a^2=[\mu^2\omega^6+(\lambda-k_2\Omega^2)^2]\frac{\Lambda^6}{\omega^2}. \label{eq.20} 
\end{eqnarray}
We determine the detuning parameter $ \sigma$ from  eq.(\ref{eq.20})
\begin{eqnarray}
 &&\sigma=(3\lambda+k_2\Omega^2)\frac{\Lambda^2}{\omega}+\frac{1}{8}(3\lambda+k_2\omega^2)\frac{a^2}{\omega}\pm\cr
&&[[\mu^2\omega^6+(\lambda-k_2\Omega^2)^2]\frac{\Lambda^6}{\omega^2a^2}-(-\frac{1}{2}+3\Omega^2\Lambda^2+
\frac{3}{8}\omega^2a^2)^2\mu^2]^{\frac{1}{2}}. \label{eq.21}
\end{eqnarray}
This equation is the frequency-response curve for superharmonic.

The peak amplitude verify the following equation
\begin{eqnarray}
 [\mu^2\omega^6+(\lambda-k_2\Omega^2)^2]\frac{\Lambda^6}{\omega^2a_p^2}-(-\frac{1}{2}+3\Omega^2\Lambda^2+
\frac{3}{8}\omega^2a_p^2)^2\mu^2=0, \label{eq.22}
\end{eqnarray}
and corresponding value of $ \sigma$ is 
\begin{eqnarray}
 \sigma_p=(3\lambda+k_2\Omega^2)\frac{\Lambda^2}{\omega}+\frac{1}{8}(3\lambda+k_2\omega^2)\frac{a_p^2}{\omega}. \label{eq.23}
\end{eqnarray}

Therefore, we noticed that the peak amplitude and frequency of this resonance are affected by third order non linearity parameters and
by the forcing amplitude but the parametric excitation term and the coefficients of the quadratics nonlinears terms does not contribute 
to this resonance at first order. 

We plot in Fig. 2 the frequency-response curve giving by  eq.(\ref{eq.20}). This curve also prouve that the amplitude of the resonance 
frequency when augment the external exciting force to which the 
order-three resonance superharmonic increases.The frequency domain where this resonance in this order appear is smaller than the case of order-two.
The peak amplitude obtain at order-two superharmonic resonance is more increased than the peak amplitude of order-three for the same resonance. 
   
\begin{figure}[htbp]
\begin{center}
 \includegraphics[width=12cm, height=6cm]{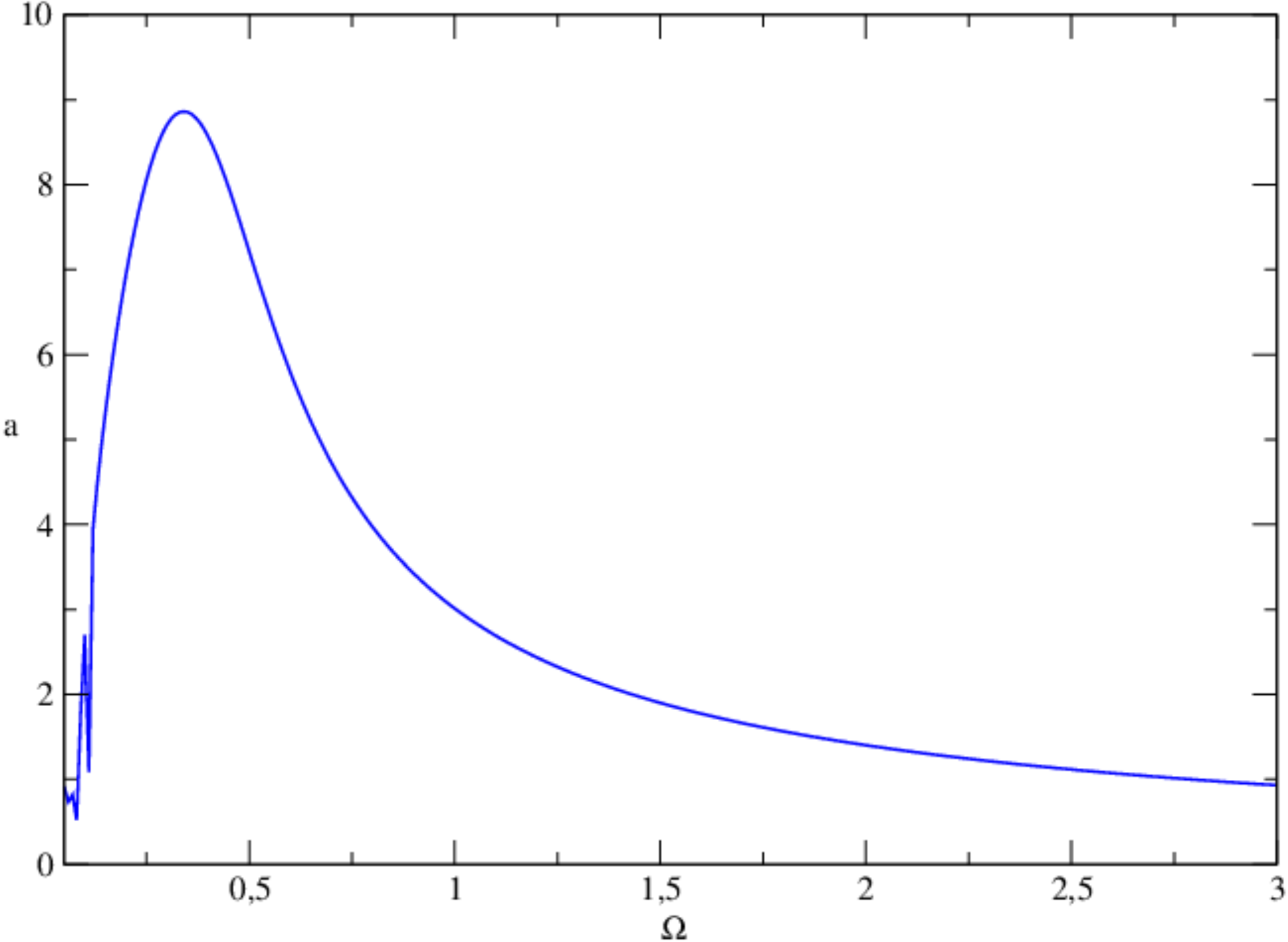}
\end{center}
\caption{superharmonic resonance in the space$(a,\Omega )$ for $\mu=0.5; \lambda=0.5;  k_2=0.5;F=0.001; \epsilon=0.01 $.}
\end{figure}

\subsection{Subharmonic resonances}

The subharmonic resonance take place if $\Omega=2\omega+\epsilon\sigma$ or $\Omega=3\omega+\epsilon\sigma$.

\subsubsection{$\Omega=2\omega+\epsilon\sigma$}
The first term and the term with $i(\Omega-\omega)T_0 $ in eq.(\ref{eq.5}) contribute to secular terms. The condition of the elimination of 
secular terms is 
\begin{eqnarray}
 &&-2i\omega A'-i\mu\omega A+3i\mu\omega^3A^2\bar{A}+6i\mu\omega\Omega^2 A\Lambda^2-3\lambda A^2\bar{A}-6\lambda A\Lambda^2-\cr
&&k_2\omega^2A^2\bar{A}-2k_2\Omega^2A\Lambda^2+(-\frac{1}{2}\alpha\bar{A}-ik_1\Omega\Lambda\bar{A}+ik_1\omega\Lambda\bar{A}-\cr
&&2\beta\omega\Omega\bar{A}\Lambda)e^{i\sigma T_1}=0, \label{eq.24}
\end{eqnarray}
We substitute the polar notation for $A$ (\ref{eq.x}) in eq.(\ref{eq.24}), and equate the real and imaginary parts, and let $\gamma=\sigma T_1-2\theta$
\begin{eqnarray}
 a'&=&(-1+6\Omega^2\Lambda^2 +\frac{3}{4}\omega^2a^2)\mu a-\frac{(\Omega+\omega)}{\omega}k_1\Lambda a\cos\gamma-\cr
&&(\frac{1}{2}\alpha+2\beta\omega\Omega\Lambda)\frac{a}{\omega}\sin\gamma, \label{eq.25}
\end{eqnarray}
and
\begin{eqnarray}
 a\gamma'&=&[\sigma-2(3\lambda+k_2\Omega^2)\frac{\Lambda^2}{\omega}-\frac{1}{4}(3\lambda+k_2\omega^2)\frac{a^2}{\omega}]a+\frac{(\Omega+\omega)}{\omega}k_1\Lambda a\sin\gamma-\cr
&&(\frac{1}{2}\alpha+2\beta\omega\Omega\Lambda)\frac{a}{\omega}\cos\gamma. \label{eq.26}
\end{eqnarray}
Seeking steady-state, we let $a'=\gamma'=0$ and we eliminate $ \gamma$ dependence to get the frequency response equation as 
\begin{eqnarray}
&& [\sigma-2(3\lambda+k_2\Omega^2)\frac{\Lambda^2}{\omega}-\frac{1}{4}(3\lambda+k_2\omega^2)\frac{a^2}{\omega}]^2a^2+(-1+6\Omega^2\Lambda^2 +\cr
&&\frac{3}{4}\omega^2a^2)^2\mu^2 a^2=[k_1^2\Lambda^2\frac{(\omega+\Omega)^2}{\omega^2}+
\frac{(\frac{1}{2}\alpha+2\beta\omega\Omega\Lambda)^2}{\omega^2}]a^2.\label{eq.27}
\end{eqnarray}

For this equation we have the trivial solution $a=0$ and another set of solutions which verify the following equation:
\begin{eqnarray}
&& [\sigma-2(3\lambda+k_2\Omega^2)\frac{\Lambda^2}{\omega}-\frac{1}{4}(3\lambda+k_2\omega^2)\frac{a^2}{\omega}]^2+\cr
&&(-1+6\Omega^2\Lambda^2 +\frac{3}{4}\omega^2a^2)^2\mu^2 =[k_1^2\Lambda^2\frac{(\omega+\Omega)^2}{\omega^2}+\frac{(\frac{1}{2}\alpha+
2\beta\omega\Omega\Lambda)^2}{\omega^2}].\label{eq.28}
\end{eqnarray}
We obtain finaly the non trivial solutions of the form
\begin{equation}
 a^2=p_1\pm(p_1^2-q_1)^{\frac{1}{2}}, \label{eq.29}
\end{equation}
where 
\begin{eqnarray}
&& p_1=\frac{4\omega(\lambda+k_2\omega^2)[\sigma-2(3\lambda+4k_2\omega^2)\frac{\Lambda^2}{\omega}]-12\mu^2\omega^4(-1+24\omega^2\Lambda^2)}{9[\mu^2\omega^6+9(3\lambda+k_2\omega^2)^2]}\\
&&q_1=\frac{1}{9[\mu^2\omega^6+9(3\lambda+k_2\omega^2)^2]}[16\omega^2[\sigma-2(3\lambda+4k_2\omega^2)\frac{\Lambda^2}{\omega}]^2+\cr
&&16\mu^2\omega^2(-1+24\omega^2\Lambda^2)^2-16k_1^2\Lambda^2(\omega+\Omega)^2-8\alpha-64\beta\omega^2\Lambda)].
\end{eqnarray}
For non trivial solutions, it follows from eq.(\ref{eq.29}) that both the radical and the first term must be positive, i.e. the 
non trivial solutions for $a$ are real only when  $p_1>0$ and $p_1^2\geq q_1$. Theses conditions imply that solutions will exist if    
\begin{eqnarray}
 4\omega(\lambda+k_2\omega^2)[\sigma-2(3\lambda+4k_2\omega^2)\frac{\Lambda^2}{\omega}]>12\mu^2\omega^4(-1+24\omega^2\Lambda^2)
\end{eqnarray}
and 
\begin{eqnarray}
 \frac{[4\omega(\lambda+k_2\omega^2)[\sigma-2(3\lambda+4k_2\omega^2)\frac{\Lambda^2}{\omega}]-12\mu^2\omega^4(-1+24\omega^2\Lambda^2)]^2}{9[\mu^2\omega^6+9(3\lambda+k_2\omega^2)^2]}
\end{eqnarray}

In Fig. 3 the frequency-response equation (\ref{eq.29}) is plotted. This curve shows that the subharmonic resonance in order-two apper and 
 the maximum amplitude corresponding to resonance increases as the resonance frequency augment remaining in the field imposed by
the conditions of occurrence of this resonance with the order.
\begin{figure}[htbp]
\begin{center}
 \includegraphics[width=12cm, height=6cm]{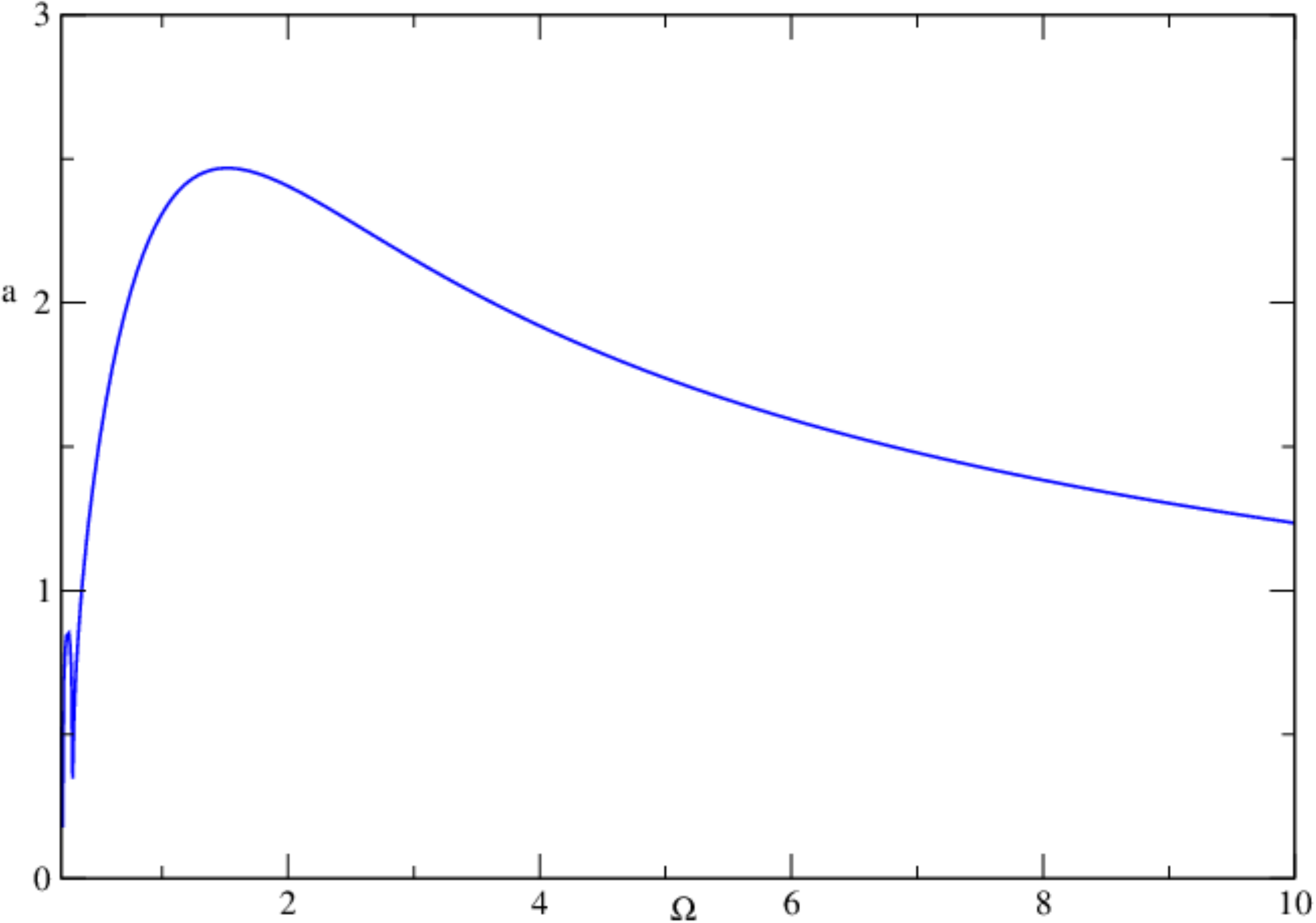}
\end{center}
\caption{subharmonic resonance in the space$(a,\Omega )$ for $\mu=0.5; \alpha=3; \lambda=-1; \beta_1=0.1; k_2=0.5; k_1=0.1; F=0.001; \epsilon=0.01 $.}
\end{figure}
\subsubsection{$\Omega=3\omega+\epsilon\sigma$}
If we insert $\Omega=3\omega+\epsilon\sigma$ in eq.(\ref{eq.5}), the solvability condition takes the form
\begin{eqnarray}
&&-2i\omega A'-i\mu\omega A+3i\mu\omega^3A^2\bar{A}+6i\mu\omega\Omega^2 A\Lambda^2-3\lambda A^2\bar{A}-\cr
&&6\lambda A\Lambda^2-k_2\omega^2A^2\bar{A}-2k_2\Omega^2A\Lambda^2+(-3i\mu\omega^2\Omega\bar{A}^2\Lambda-\cr
&&3\lambda\bar{A}^2\Lambda+k_2\omega^2\bar{A}^2\Lambda-2k_2\omega\Omega\bar{A}^2\Lambda)e^{i\sigma T_1}, \label{eq.30} 
\end{eqnarray}
Reused the rule in the undo case, and let $\gamma=\sigma T_1-3\theta$, put $a'=\gamma'=0$ and eliminate $ \gamma$, the frequency-response
equation is
\begin{eqnarray}
 &&[\sigma-3(3\lambda+k_2\Omega^2)\frac{\Lambda^2}{\omega}-\frac{3}{8}(3\lambda+k_2\omega^2)\frac{a^2}{\omega}]^2a^2+\cr
&&[(-\frac{3}{2}+9\Omega^2\Lambda)\mu+\frac{9}{8}\omega^2a^2]^2a^2=\cr
&&\frac{9}{16}[9\mu^2\omega^4\Omega^2-(3\lambda-k_2\omega^2+2k_2\omega\Omega)^2]\frac{\Lambda^2a^4}{\omega^2}. \label{eq.31}
\end{eqnarray}
The solutions of equation (\ref{eq.31}) are either $a=0$ or 
\begin{equation}
 a^2=p\pm(p^2-q)^{\frac{1}{2}}, \label{eq.32}
\end{equation}
 where 
\begin{eqnarray}
 &&p=\frac{1}{9[\mu^2\omega^6+9(3\lambda+k_2\omega^2)^2]}\times\cr
&&[-72\omega^4(-\frac{3}{2}+9\Omega^2\Lambda)\mu+24\omega(3\lambda+k_2\omega^2)[\sigma-3(3\lambda+k_2\Omega^2)\frac{\Lambda^2}{\omega}]+\cr
&&18[9\mu^2\omega^4\Omega^2-(3\lambda-k_2\omega^2+2k_2\omega\Omega)^2
]\Lambda^2]
\end{eqnarray}
\begin{eqnarray}
q=\frac{64\omega^2(-\frac{3}{2}+9\Omega^2\Lambda)^2\mu^2+64\omega[\sigma-3(3\lambda+k_2\Omega^2\Lambda^2)]^2}{9[\mu^2\omega^6+9(3\lambda+k_2\omega^2)^2]}.
\end{eqnarray}

Since $q$ is always positive, we need $p>0 $ and $ p^2\geq q$. This requires that
\begin{eqnarray}
&&24\omega(3\lambda+k_2\omega^2)[\sigma-3(3\lambda+k_2\Omega^2)\frac{\Lambda^2}{\omega}]+\cr
&&18[9\mu^2\omega^4\Omega^2-(3\lambda-k_2\omega^2+2k_2\omega\Omega)^2]\Lambda^2]>72\omega^4(-\frac{3}{2}+9\Omega^2\Lambda)\mu
\end{eqnarray}
and
\begin{eqnarray}
&&[-72\omega^4(-\frac{3}{2}+9\Omega^2\Lambda)\mu+24\omega(3\lambda+k_2\omega^2)[\sigma-3(3\lambda+k_2\Omega^2)\frac{\Lambda^2}{\omega}]+\cr
&&18[9\mu^2\omega^4\Omega^2-(3\lambda-k_2\omega^2+2k_2\omega\Omega)^2]\Lambda^2]^2\geq\cr
&&9[64\omega^2(-\frac{3}{2}+9\Omega^2\Lambda)^2\mu^2+64\omega[\sigma-3(3\lambda+k_2\Omega^2\Lambda^2)]^2]\times\cr
&&[\mu^2\omega^6+9(3\lambda+k_2\omega^2)^2]
\end{eqnarray}

We simulate eq.(\ref{eq.32}) (see Fig. 4) and we  note the same comments as the case of the subharmonic of order-two but the resonance amplitudes for resonance are
 significant in this case i.e. the order-three. 
\begin{figure}[htbp]
\begin{center}
 \includegraphics[width=12cm, height=6cm]{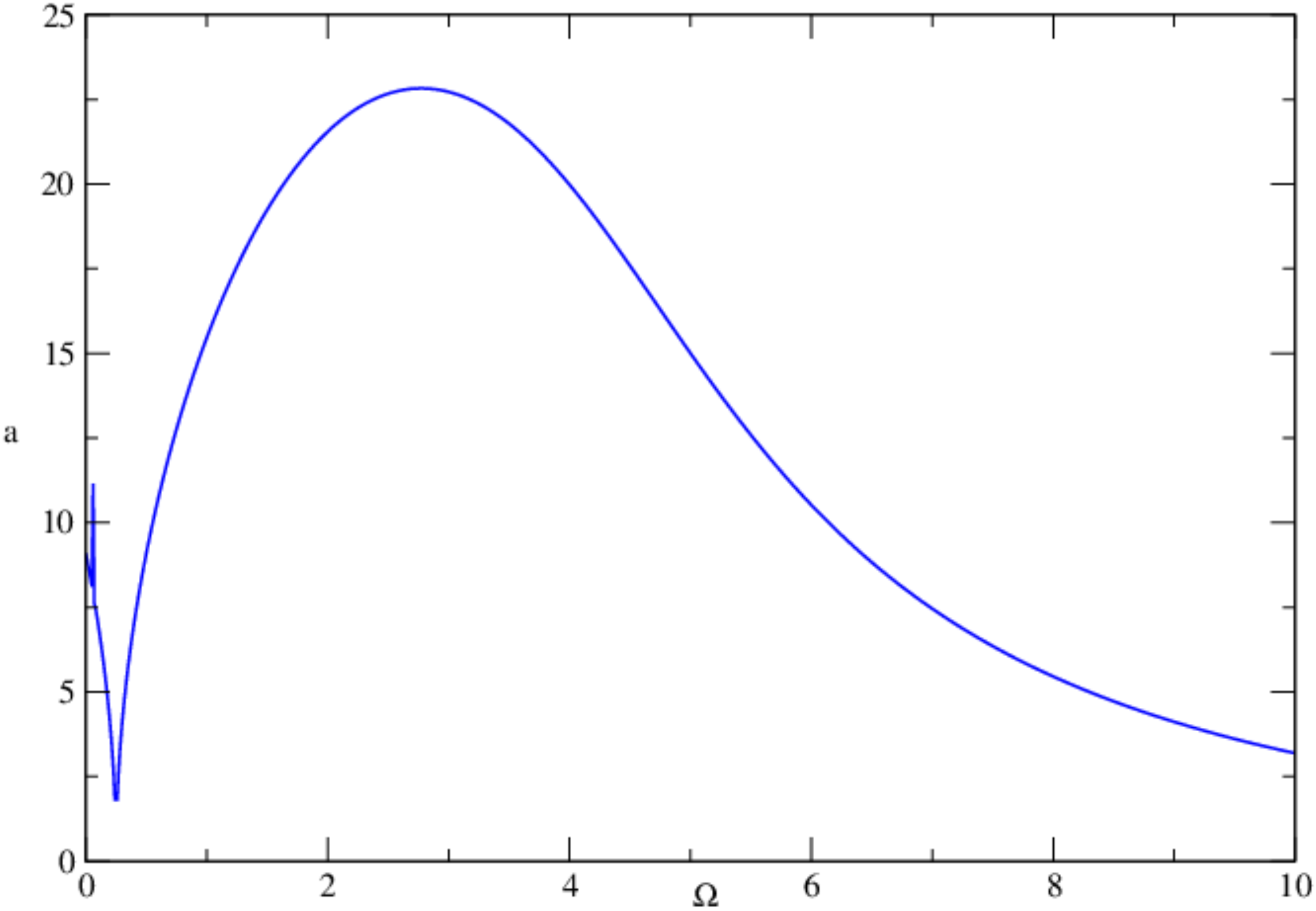}
\end{center}
\caption{subharmonic resonance in the space$(a,\Omega )$ for $\mu=0.5; \alpha=3; \lambda=-1; \beta_1=0.1; k_2=0.5; k_1=0.1;F=0.001; \epsilon=0.01 $.}
\end{figure}
\subsection{primary resonance}
In this state, we put that $F=\epsilon F$.
 The closeness between both internal and external frequencies is given by 
 $\Omega = 1+\epsilon \sigma$. In these conditions after some algebaic manipulations, we obtain 

\begin{eqnarray}
 D_0^2x_1+\omega^2x_1&=& [-2i\omega A'-i\mu\omega A+3i\mu\omega^3 A^2\bar{A}-k_2\omega^2A^2\bar{A}-\cr
&&3\lambda A^2A+\frac{1}{2}F\exp{i\sigma T_1}]e^{i\omega T_0}+cc+NST,\label{eq.33}
\end{eqnarray}

equating resonant terms at $0$
 from Eq.(\ref{eq.33}), we obtain:
\begin{eqnarray}
 &&-2i\omega A'-i\mu\omega A+3i\mu\omega^3 A^2\bar{A}-k_2\omega^2A^2\bar{A}-\cr
&&3\lambda A^2A+\frac{1}{2}F\exp{i\sigma T_1}=0. \label{eq.34}
\end{eqnarray}

Afer the same algebraic manipulations in other resonant states, the amplitude of oscillations of primary resonant
 states  is governed by the  following nonlinear algebraic equation.

\begin{eqnarray} 
 (-\frac{1}{2}\mu\omega a+\frac{3}{8}\mu\omega^3a^3)^2+(\omega\sigma a-\frac{1}{2}k_2\omega^2 a^3-\frac{3}{8}\lambda a^3)^2=\frac{F^2}{4}.\label{eq.35}
\end{eqnarray}
\section{Effects of  parameters on  resonant states }
In this section, we study separately the effect of each parameter on each  resonance of the oscillator i.e. the effect of different nonlinearity
 parameter and the amplitude of the exciting force on the differents resonances which appear for this modified Rayleigh-Duffing oscillator. 

\subsection{Effects of parameters on  superharmonic resonance}
The first three figures respectively  show the effects of the cubic nonlinear parameters for the order-three of the superharmonic resonance and 
the following six figures respectively show the effects of the $\alpha, \beta, k_1, \mu, \lambda$  and $k_2$  for the two levels of the superharmonic resonance
The superharmonic response of order 1/2 involves interaction between the parametric excitation and both the nonlinear parameter and the direct 
excitation. In fact, if the nonlinearity is not present, this resonance persists.

Figure 5 shows that for the first  resonance, the  parameter $\mu$ has no effect. For the two and third superharmonic resonance of order-two that appears,
note that most $\mu$ parameter increases the higher the frequency of occurrence and the maximum value of the amplitude of the oscillator of this 
resonance decreases and the width of the curve resonance is less.

From Figure 6, we note that the same is the case of $\mu$ parameter except that here the sulfide and as $\lambda$ increases the resonance amplitude decreases, 
the frequency of occurrence of superharmonic resonance of order two the third time and increasing the resonance curve becomes broader.

Analysis curve of the figure 7, we find that $k_2$ and $\mu$ have virtually the same effect on the  supharmonic resonance order-two.

Figure 8 shows that the amplitude of the exciting force does not have a great effect on the superharmonic resonance of order-two. Figures 9 and 10
 show that the  parameters $\beta$ and $k_1$  when do increase slightly increase the amplitude of the resonance at its second appearance without 
changing the frequency with which it entry appears.

Figures 13, 14 and 15 respectively illustrate the effects of parameters $\mu$, $ \lambda$  and $k_2 $ on superharmonic resonance of order-three.
We note that these parameters have almost the same effects for the order of the resonance as in the case of the order-two.

Figures 11 and 12 show respectively the detuning parameter and the excitation force amplitude effects on superharmonic resonances order-two responses curves 
and figures 16 and 17 illustrate the effects of the parameters on superharmonic resonances order-three responses curves. Theses prove that these two parameters also
 affect severaly the resonance response curve.
 
\begin{figure}[htbp]
\begin{center}
 \includegraphics[width=12cm, height=6cm]{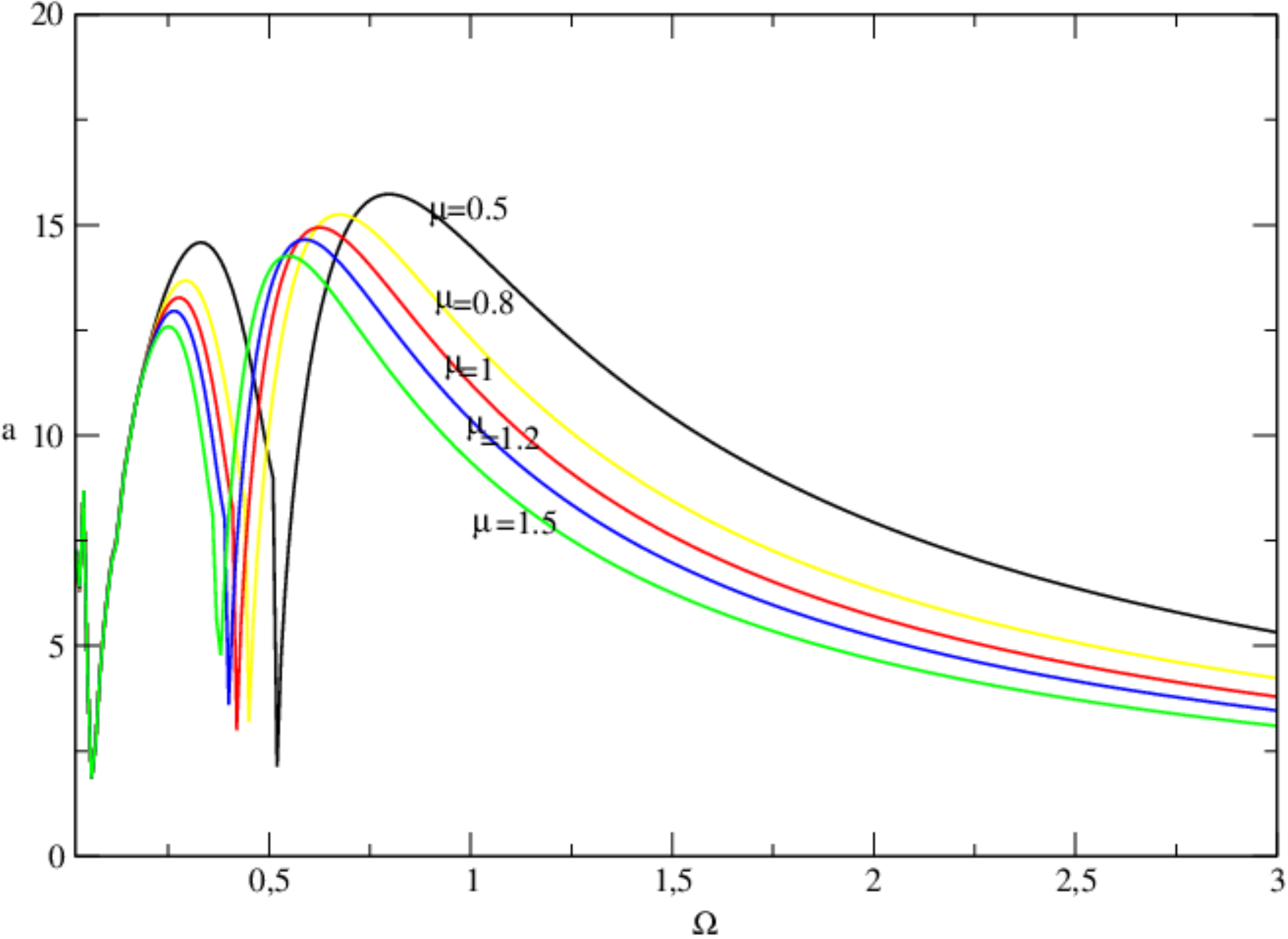}
\end{center}
\caption{Effects of $\mu$ on the frequency-response curves of the order-two superharmonic resonance with the parameters of figure 1.}
\end{figure}

\begin{figure}[htbp]
\begin{center}
 \includegraphics[width=12cm, height=6cm]{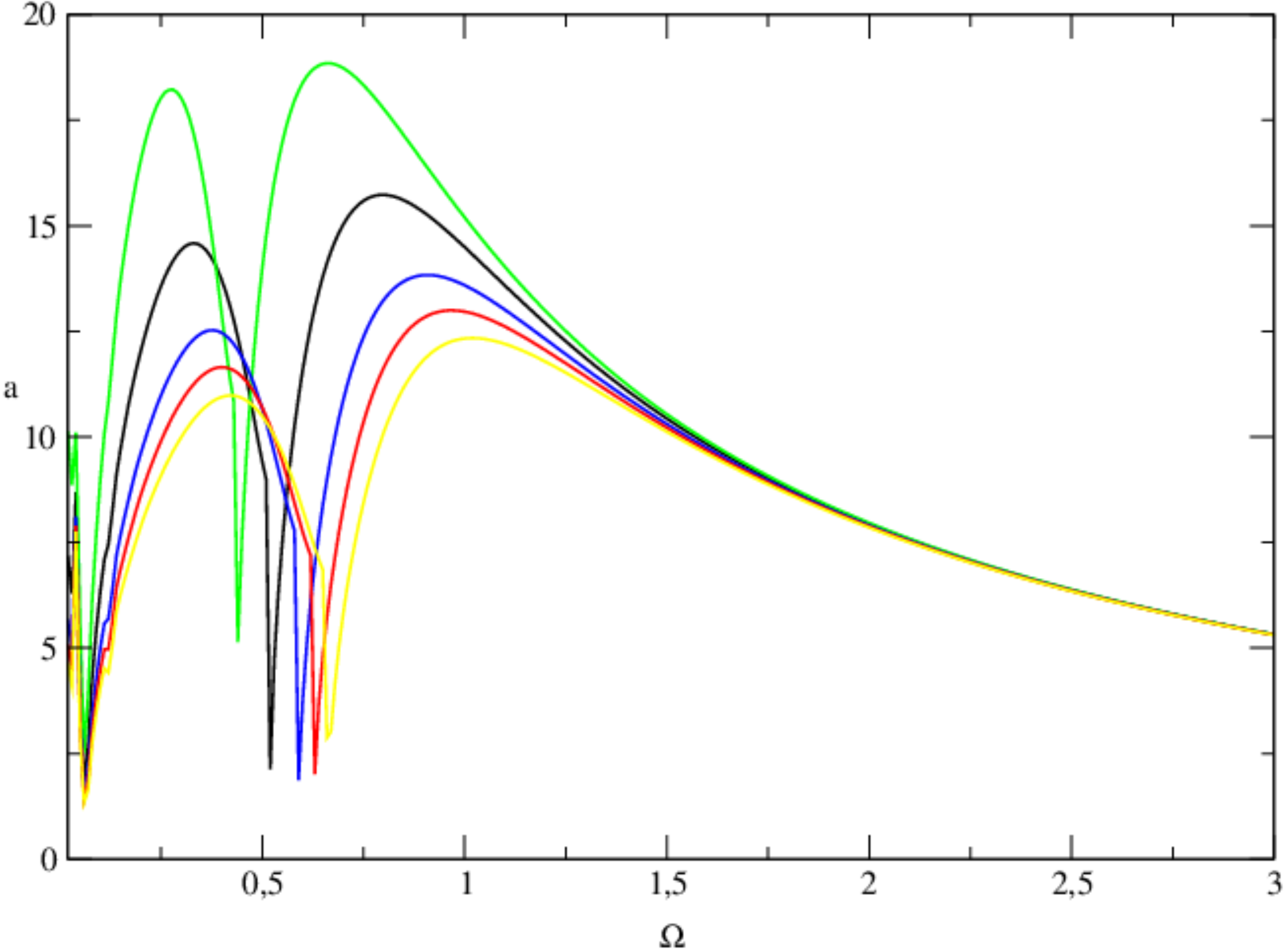}
\end{center}
\caption{Effects of $\lambda$ on the frequency-response curves of the order-two superharmonic resonance with the parameters of figure 1.}
\end{figure}

\begin{figure}[htbp]
\begin{center}
 \includegraphics[width=12cm, height=6cm]{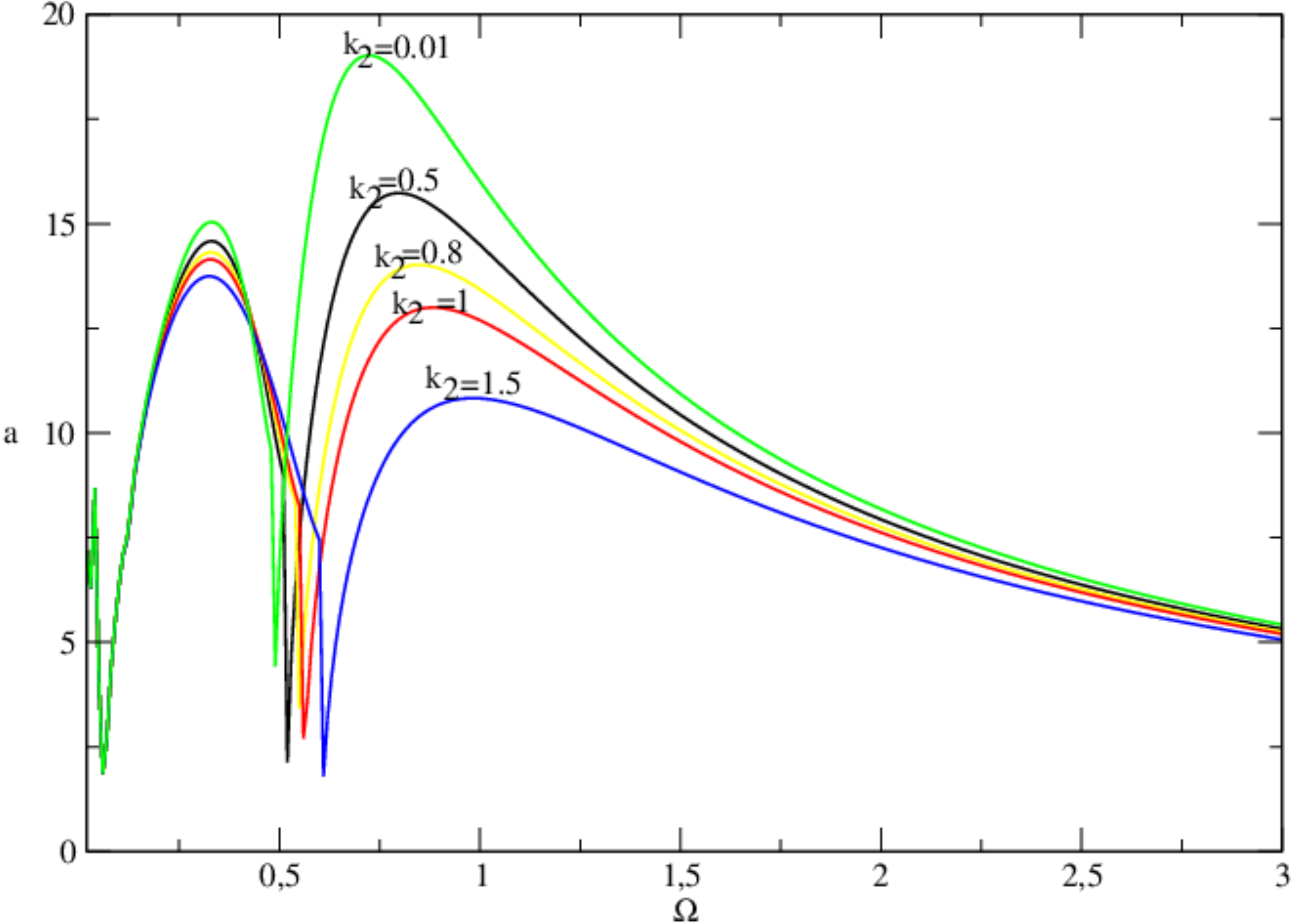}
\end{center}
\caption{Effects of $k_2$ on the frequency-response curves of the order-two superharmonic resonance with the parameters of figure 1.}
\end{figure}

\begin{figure}[htbp]
\begin{center}
 \includegraphics[width=12cm, height=6cm]{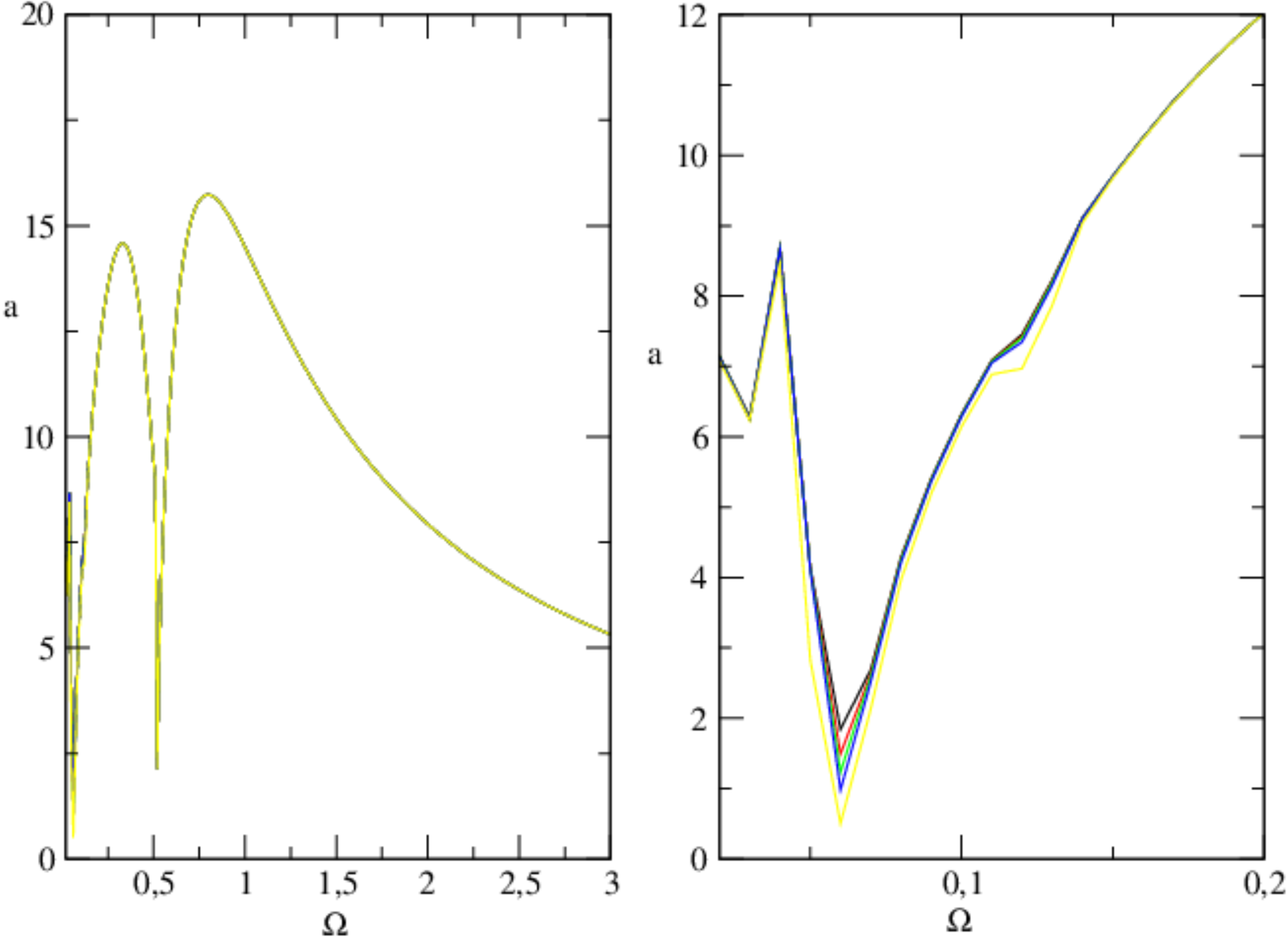}
\end{center}
\caption{Effects of $\alpha$ on the frequency-response curves of the order-two superharmonic resonance with the parameters of figure 1.}
\end{figure}

\begin{figure}[htbp]
\begin{center}
 \includegraphics[width=12cm, height=6cm]{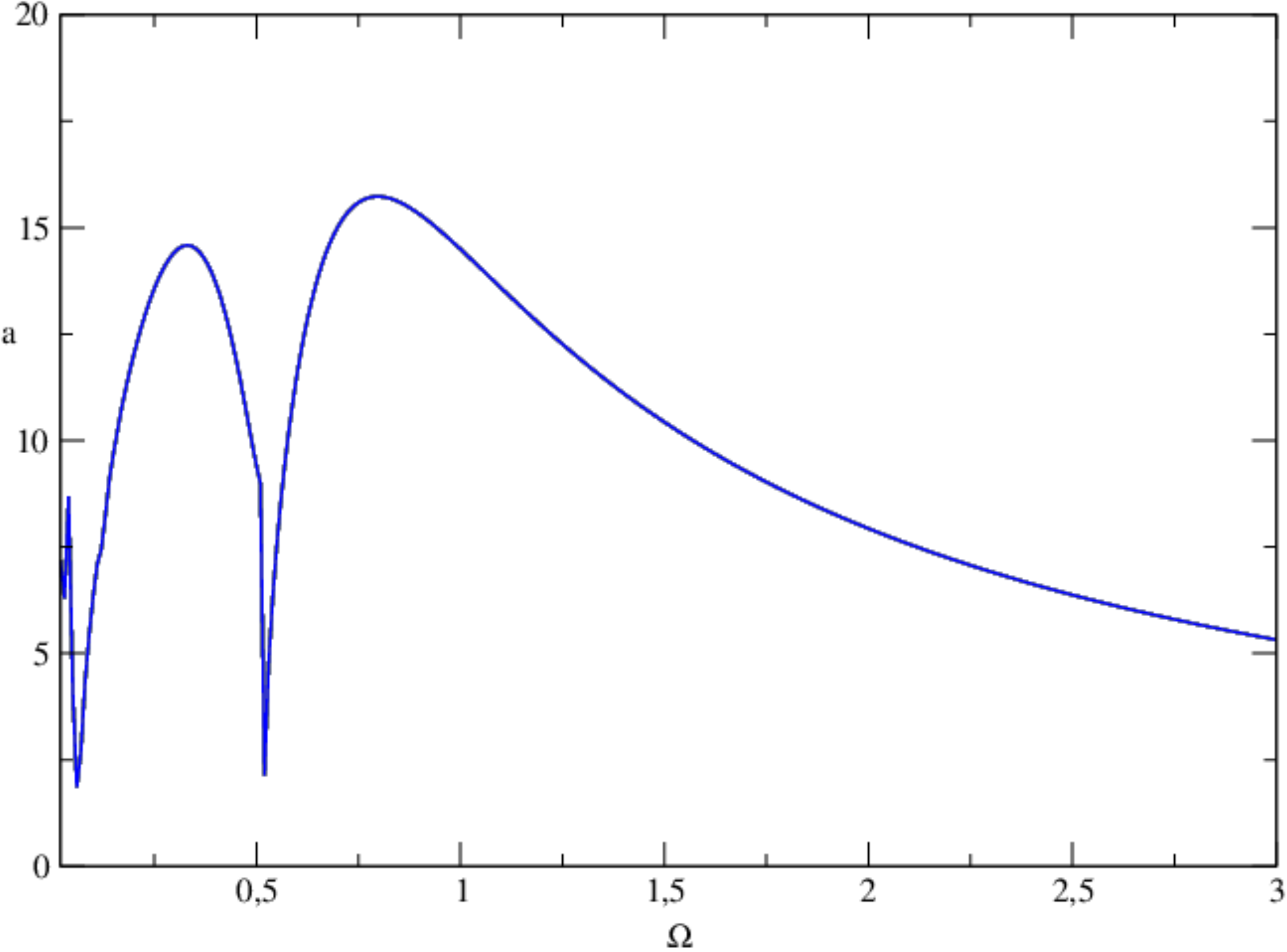}
\end{center}
\caption{Effects of $\beta$ on the frequency-response curves of the order-two superharmonic resonance with the parameters of figure 1.}
\end{figure}

\begin{figure}[htbp]
\begin{center}
 \includegraphics[width=12cm, height=6cm]{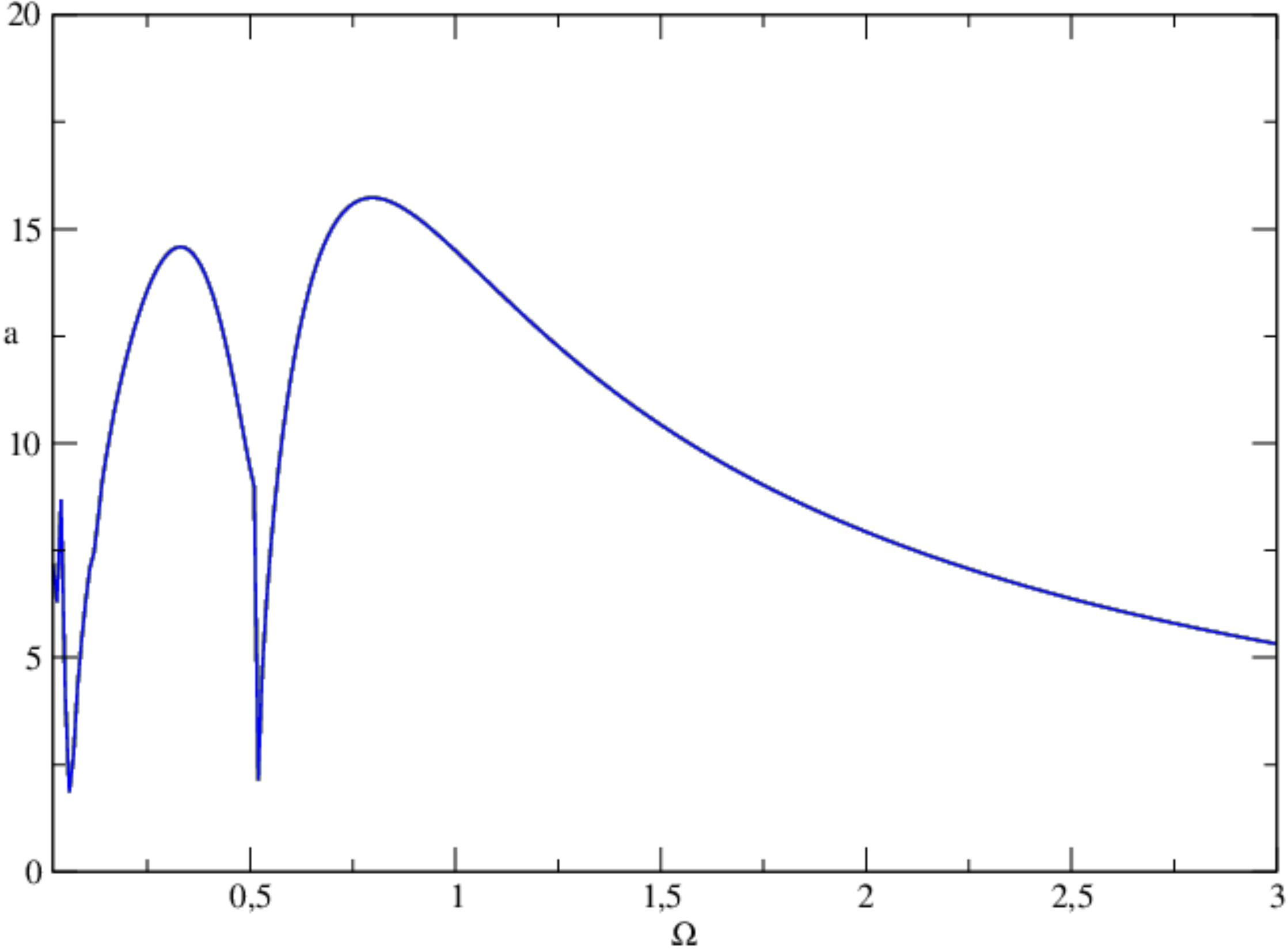}
\end{center}
\caption{Effects of $k_1$ on the frequency-response curves of the order-two superharmonic resonance with the parameters of figure 1.}
\end{figure}

\begin{figure}[htbp]
\begin{center}
 \includegraphics[width=12cm, height=6cm]{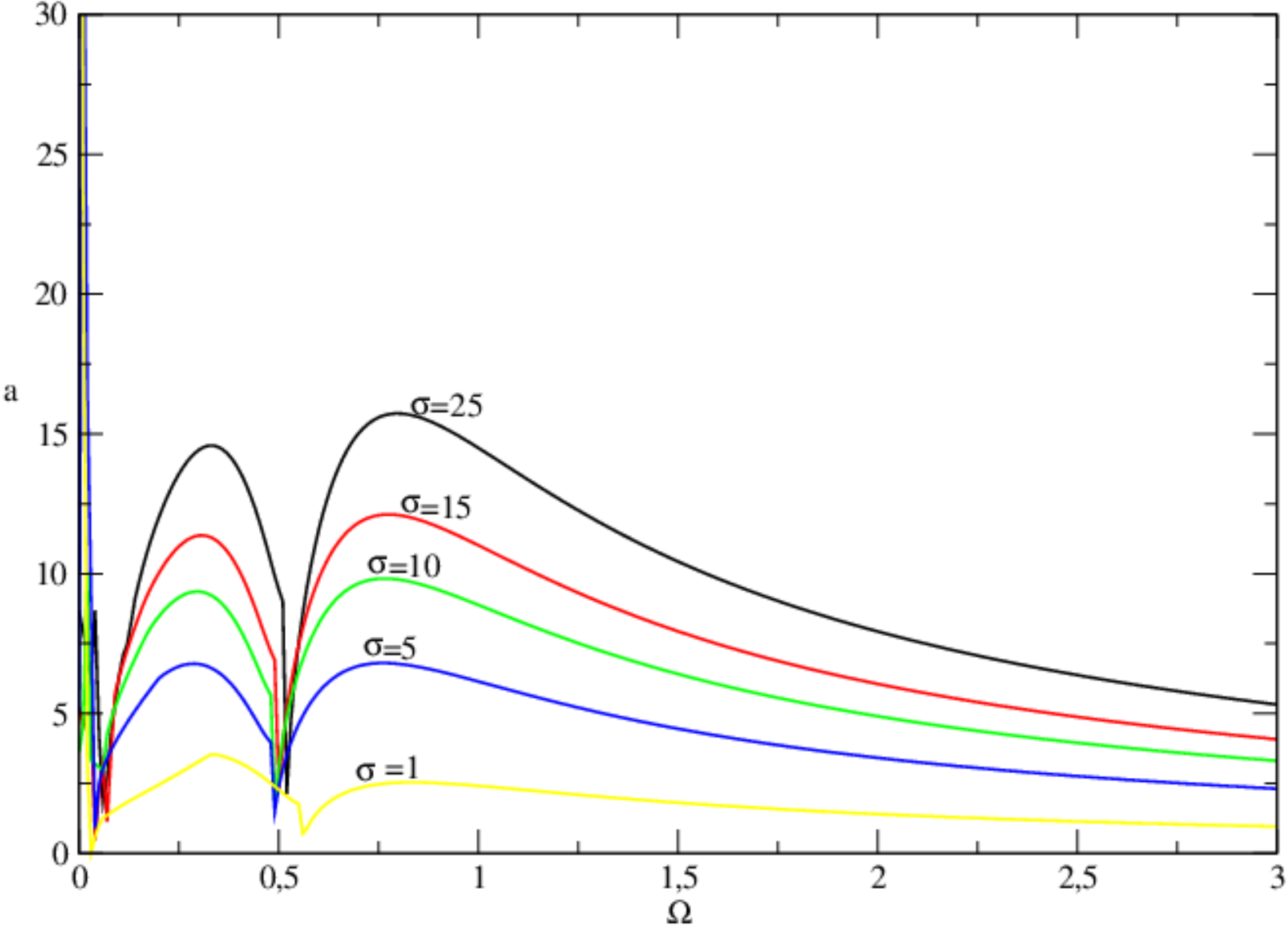}
\end{center}
\caption{Effects of $\sigma$ on the frequency-response curves of the order-two superharmonic resonance with the parameters of figure 1.}
\end{figure}

\begin{figure}[htbp]
\begin{center}
 \includegraphics[width=12cm, height=6cm]{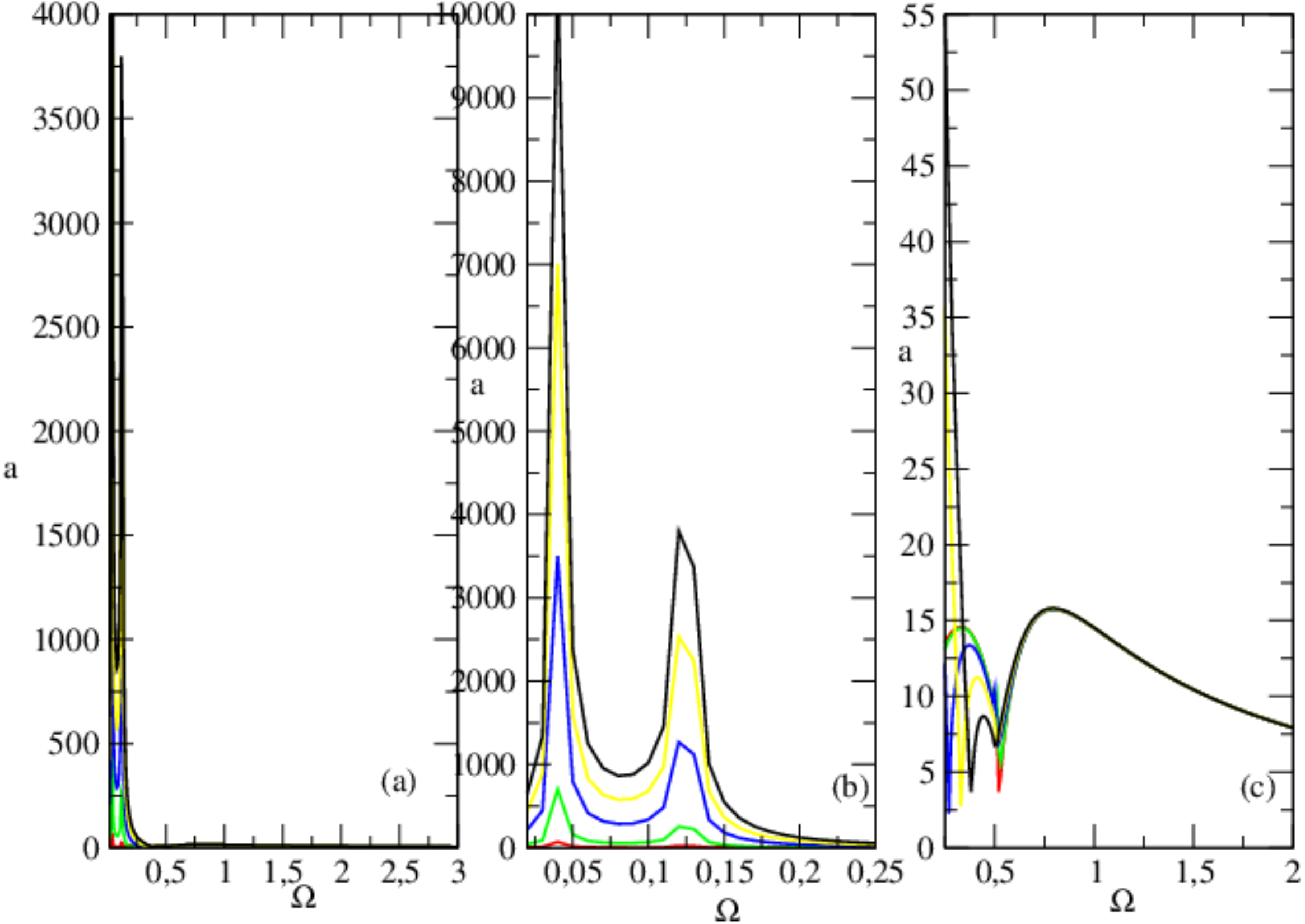}
\end{center}
\caption{Effects of $F$ on the frequency-response curves of the order-two superharmonic resonance with the parameters of figure 1.}
\end{figure}

\begin{figure}[htbp]
\begin{center}
 \includegraphics[width=12cm, height=6cm]{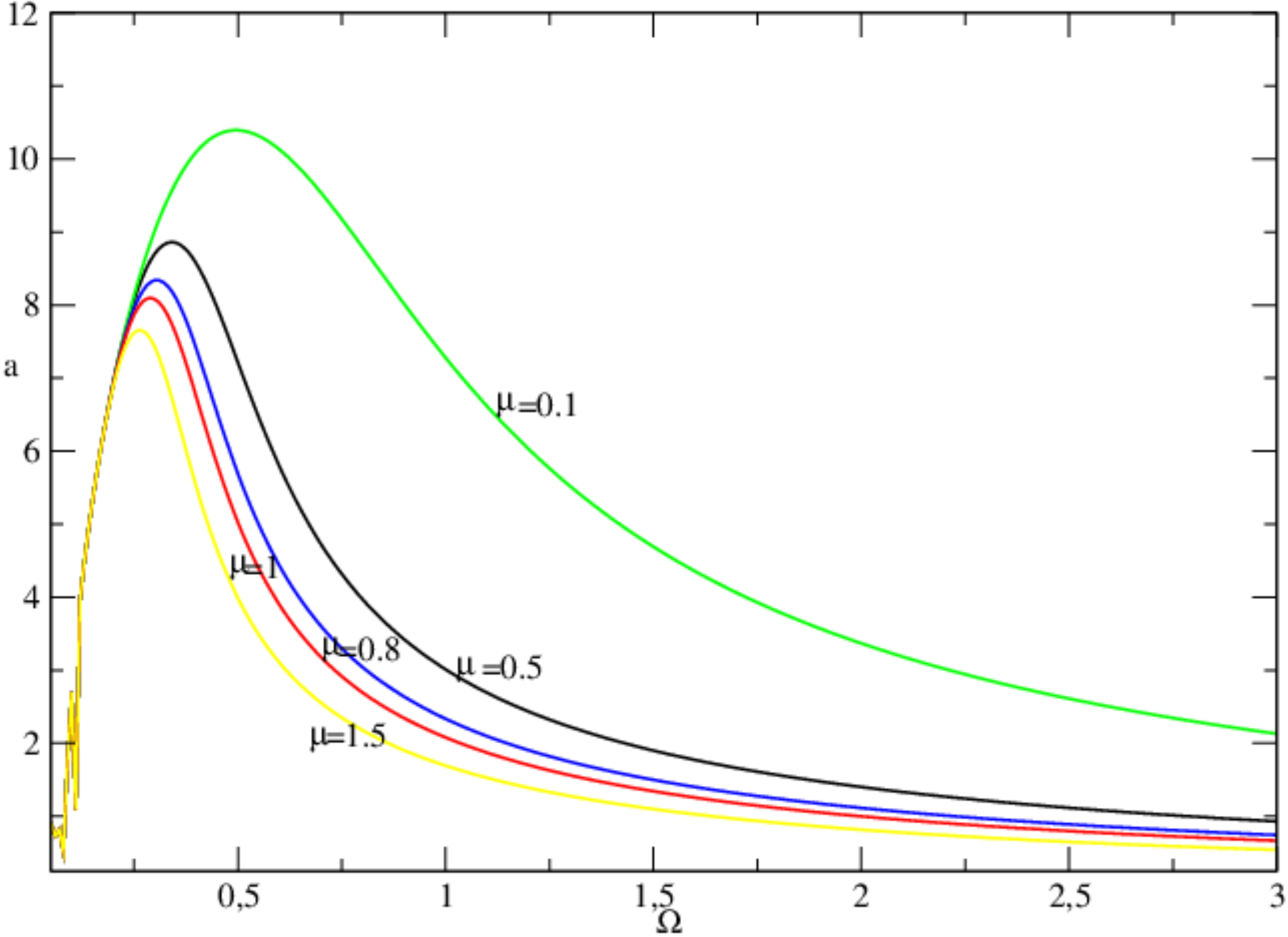}
\end{center}
\caption{Effects of $\mu$ on the frequency-response curves of the order-three superharmonic resonance with the parameters of figure 2.}
\end{figure}

\begin{figure}[htbp]
\begin{center}
 \includegraphics[width=12cm, height=6cm]{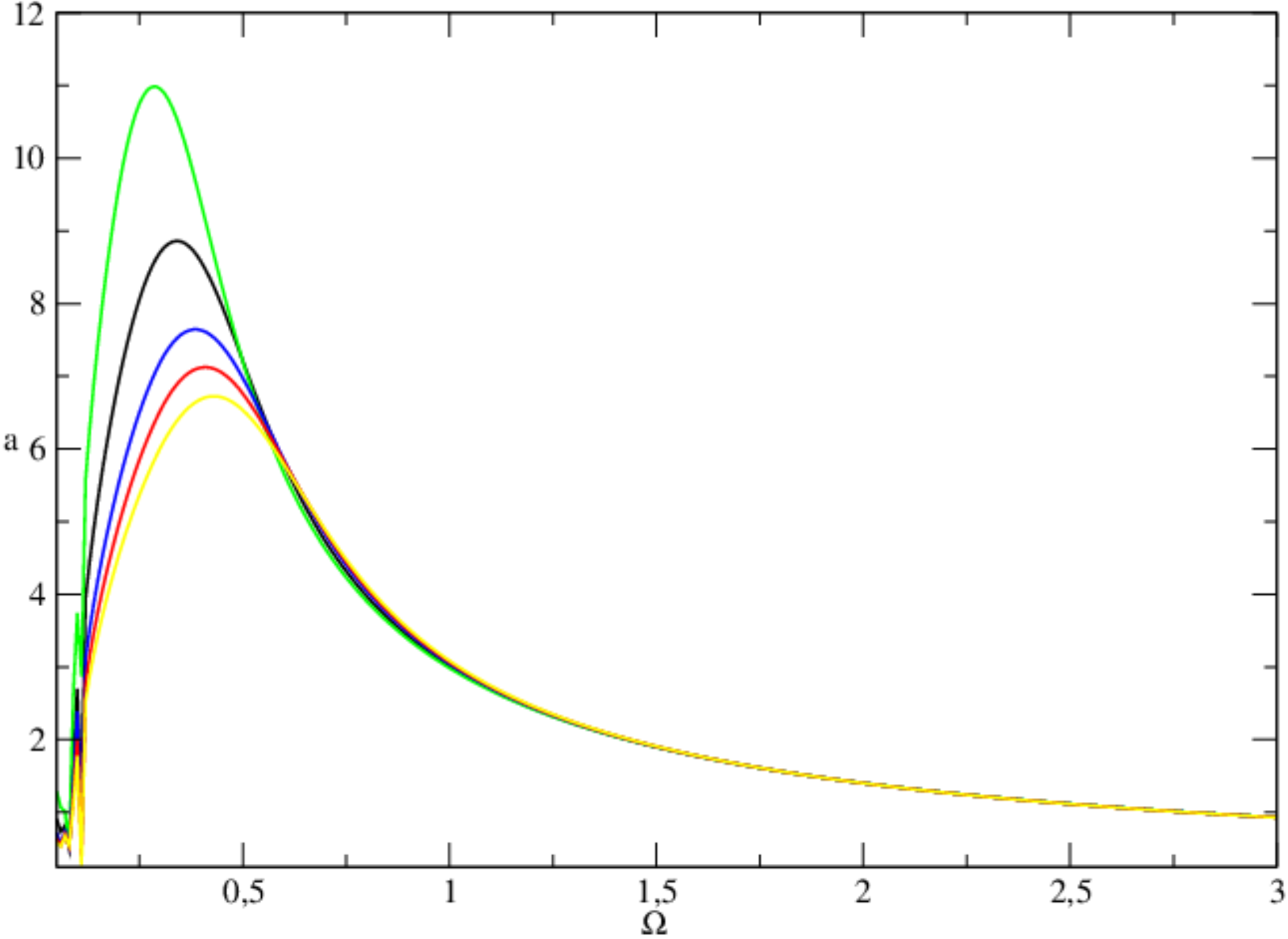}
\end{center}
\caption{Effects of $\lambda$ on the frequency-response curves of the order-three superharmonic resonance with the parameters of figure 2.}
\end{figure}

\begin{figure}[htbp]
\begin{center}
 \includegraphics[width=12cm, height=6cm]{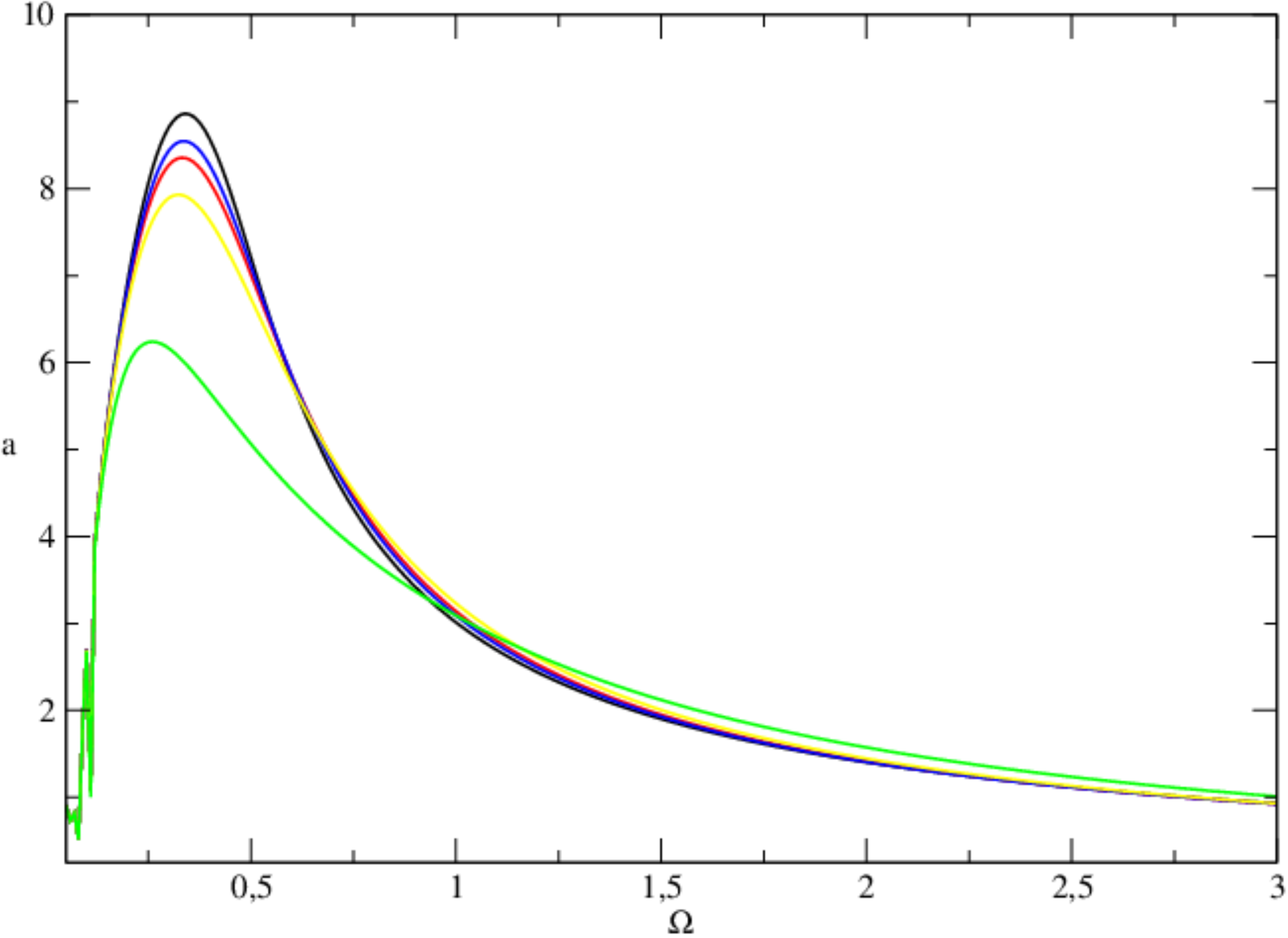}
\end{center}
\caption{Effects of $k_2$ on the frequency-response curves of the order-three superharmonic resonance with the parameters of figure 2.}
\end{figure}

\begin{figure}[htbp]
\begin{center}
 \includegraphics[width=12cm, height=6cm]{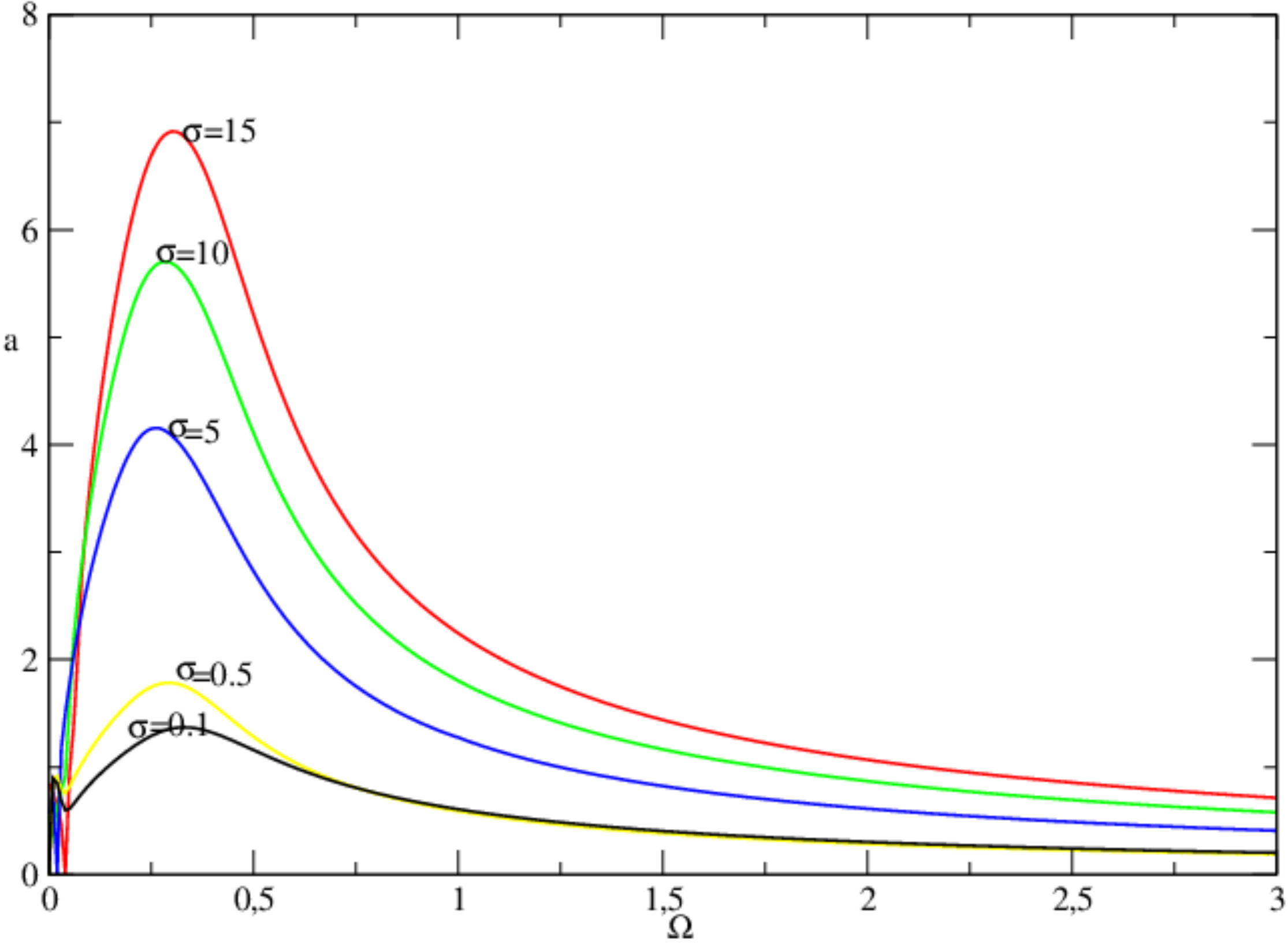}
\end{center}
\caption{Effects of $\sigma$ on the frequency-response curves of the order-three superharmonic resonance with the parameters of figure 2.}
\end{figure}

\begin{figure}[htbp]
\begin{center}
 \includegraphics[width=12cm, height=6cm]{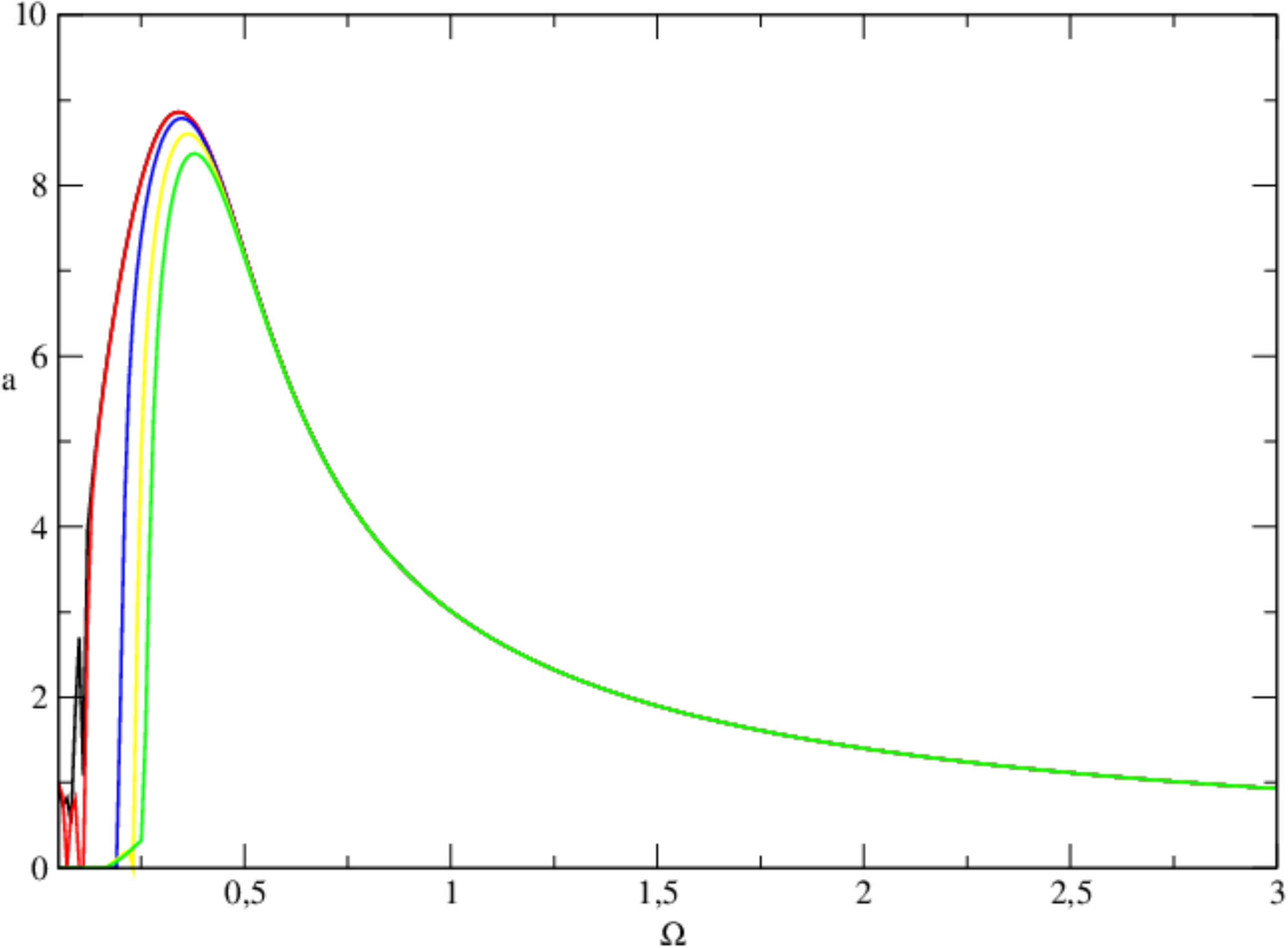}
\end{center}
\caption{Effects of $F$ on the frequency-response curves of the order-three superharmonic resonance with the parameters of figure 2.}
\end{figure}

\newpage
\subsection{Effects of  parameters on  subharmonic resonance}
In this part, we are found the effects of the same parameter. For the subharmonic resonance of order-two $ \mu$ (Figure 20) and $k_2$ (figure 19)
 have exactly the same effects as in the case of superharmonic resonance of the same order. $\lambda$ (Figure18) in turn increases the value of the resonance
 amplitude of the oscillator whenever the two-order resonance appears by increasing its frequency of occurrence for the second time.
 More $\lambda$ is large, the resonance frequency is high, which makes the resonance disappear.  In the case of subharmonic resonance for this 
oscillator, when the excitation amplitude (figure 21) and $k_1$ parameter (figure22) increase the resonance amplitude increases 
to its first appearance in keeping its frequency. We notice that the beta parameter has no effect on the behavior of the oscillator in this resonance.
In the case of subharmonic resonance of order-three (Figures 23,24 and 25), the $k_2$ parameter has exactly the same effect as if the superharmonic resonance of the same
 order. $\lambda$ is the behavior of the system in exactly the same way as the case of the order-two of this resonance.
 The parameter $\mu$ is practically no effect on the resonance that order.

 \begin{figure}[htbp]
 \begin{center}
  \includegraphics[width=12cm, height=8cm]{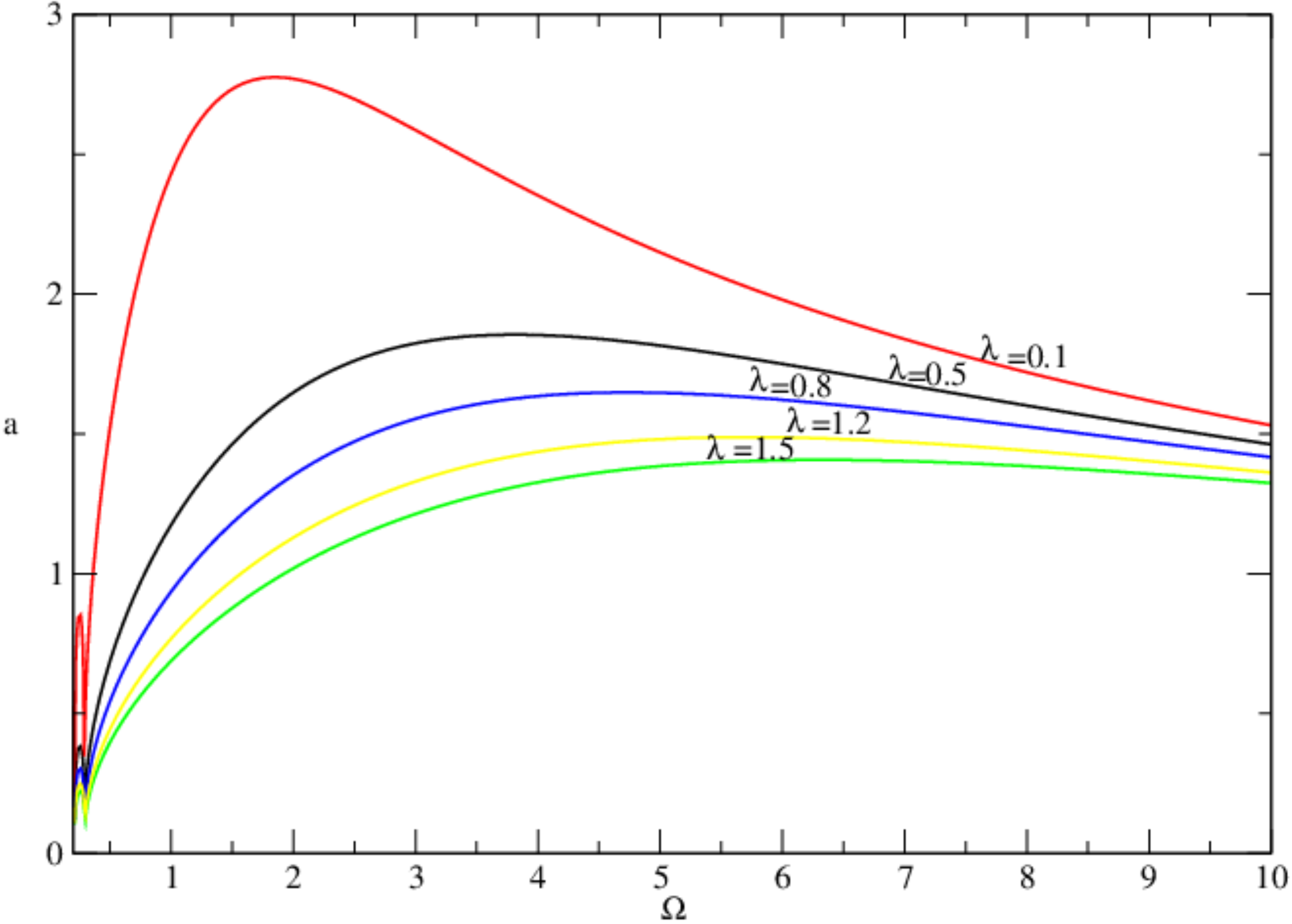}
 \end{center}
 \caption{Effects of $\lambda$ on the frequency-response curves of the order-two subharmonic resonance with the parameters of figure 3.}
 \end{figure}
 
 \begin{figure}[htbp]
 \begin{center}
  \includegraphics[width=10cm, height=6cm]{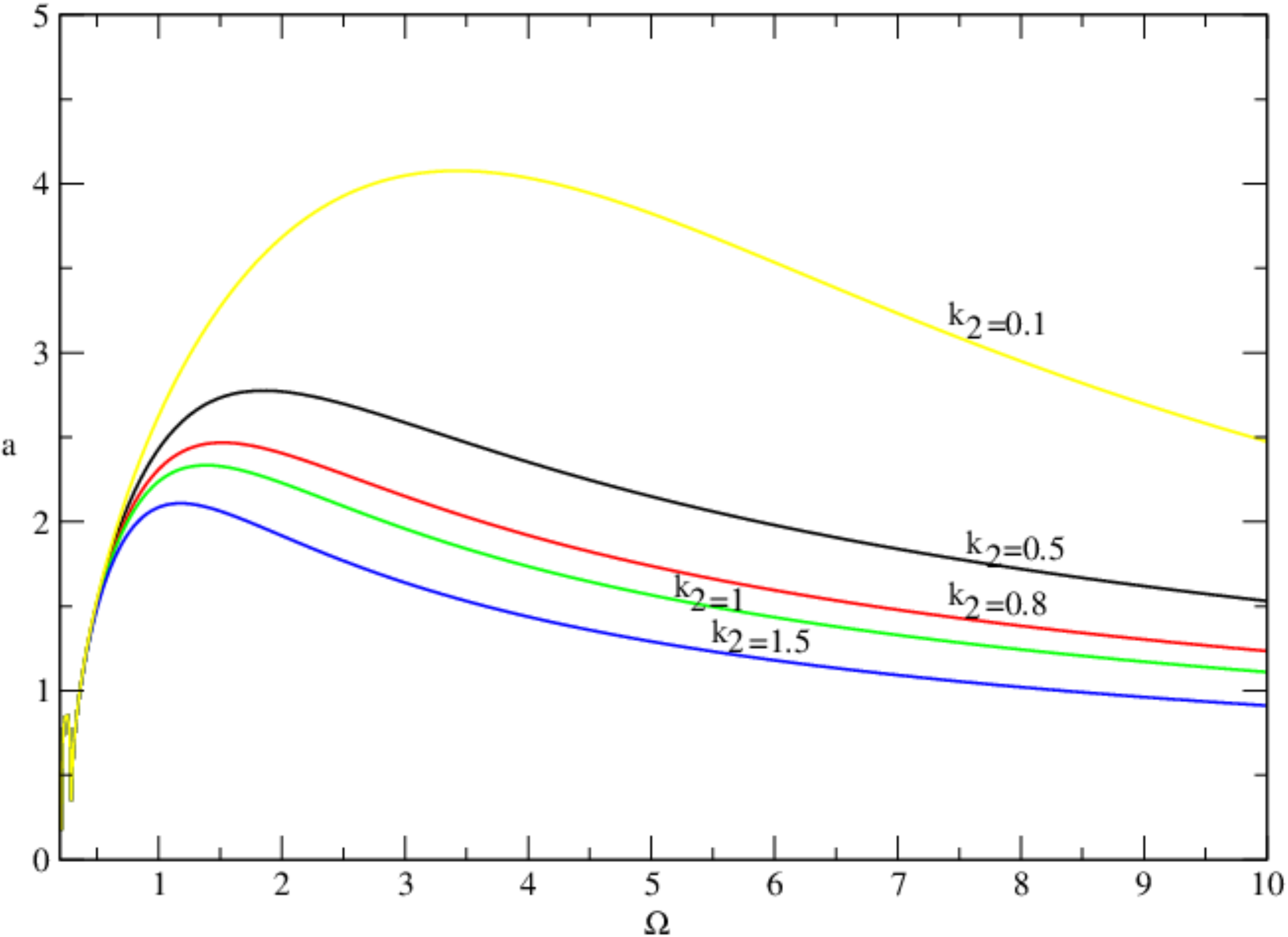}
 \end{center}
 \caption{Effects of $k_2$ on the frequency-response curves of the order-two subharmonic resonance with the parameters of figure 3.}
 \end{figure}

\begin{figure}[htbp]
\begin{center}
 \includegraphics[width=12cm, height=6cm]{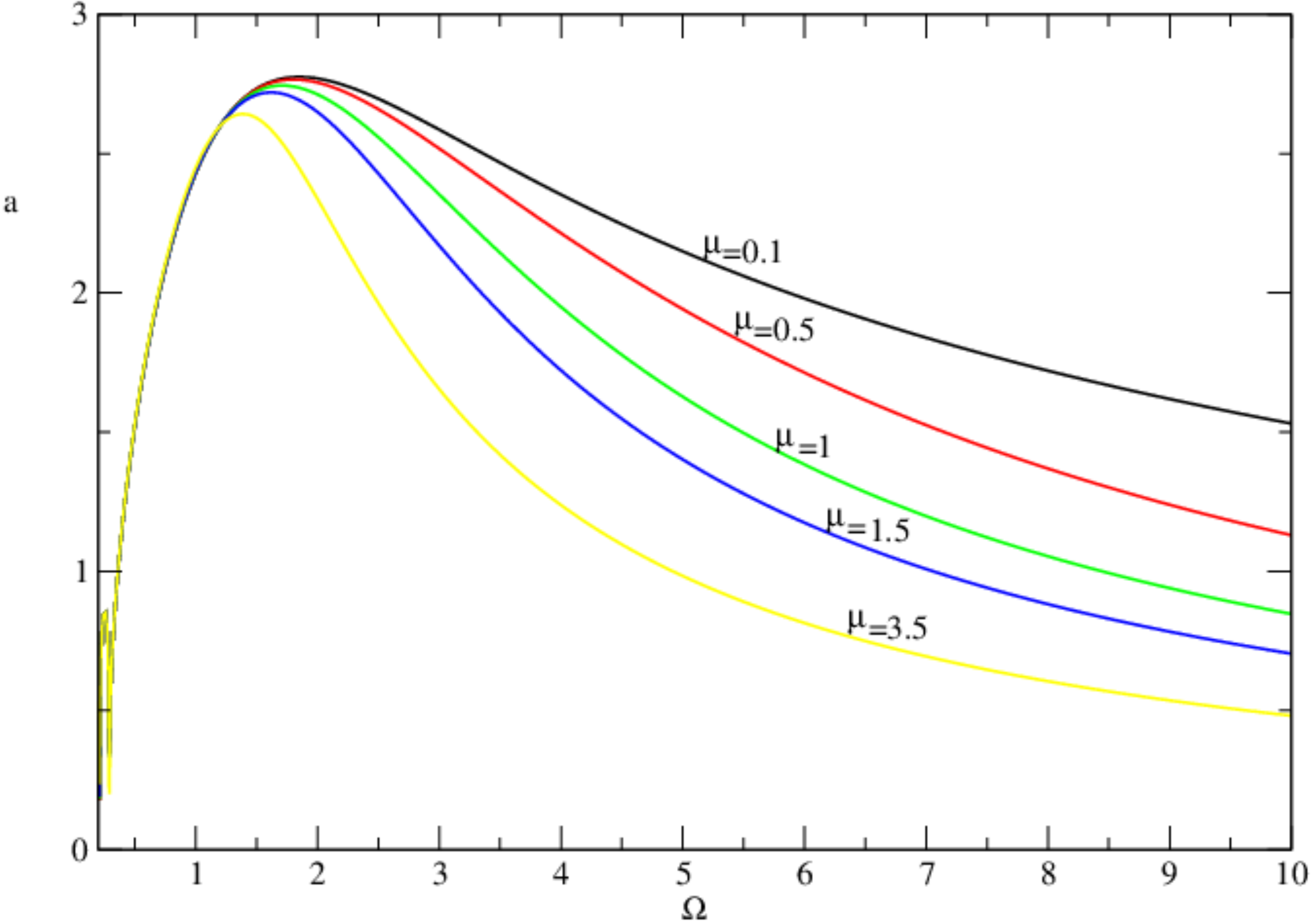}
\end{center}
\caption{Effects of $\mu$ on the frequency-response curves of the order-two subharmonic resonance with the parameters of figure 3}
\end{figure}

\begin{figure}[htbp]
\begin{center}
 \includegraphics[width=12cm, height=6cm]{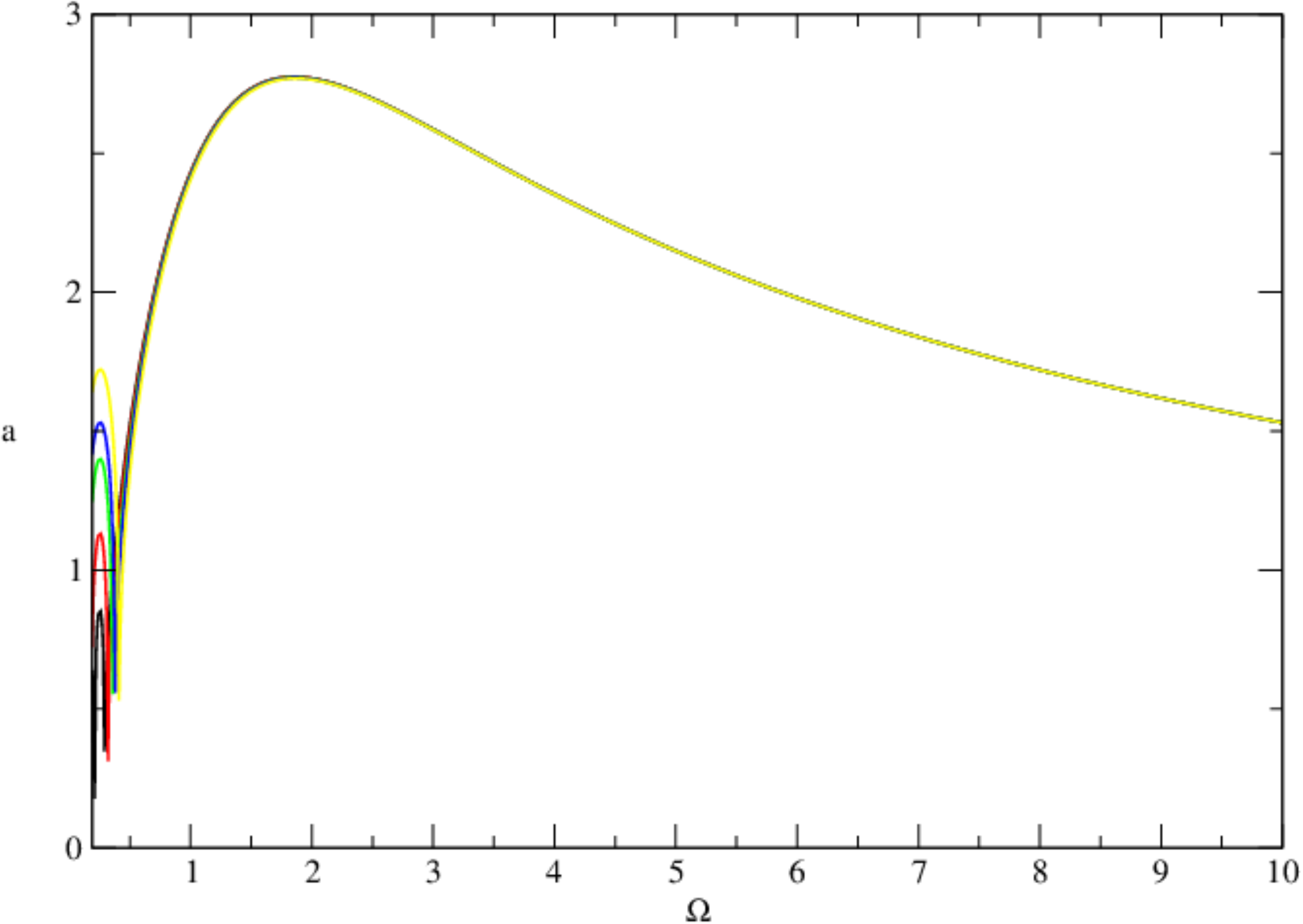}
\end{center}
\caption{Effects of $\alpha$ on the frequency-response curves of the order-two subharmonic resonance with the parameters of figure 3.}
\end{figure}

\begin{figure}[htbp]
\begin{center}
 \includegraphics[width=12cm, height=6cm]{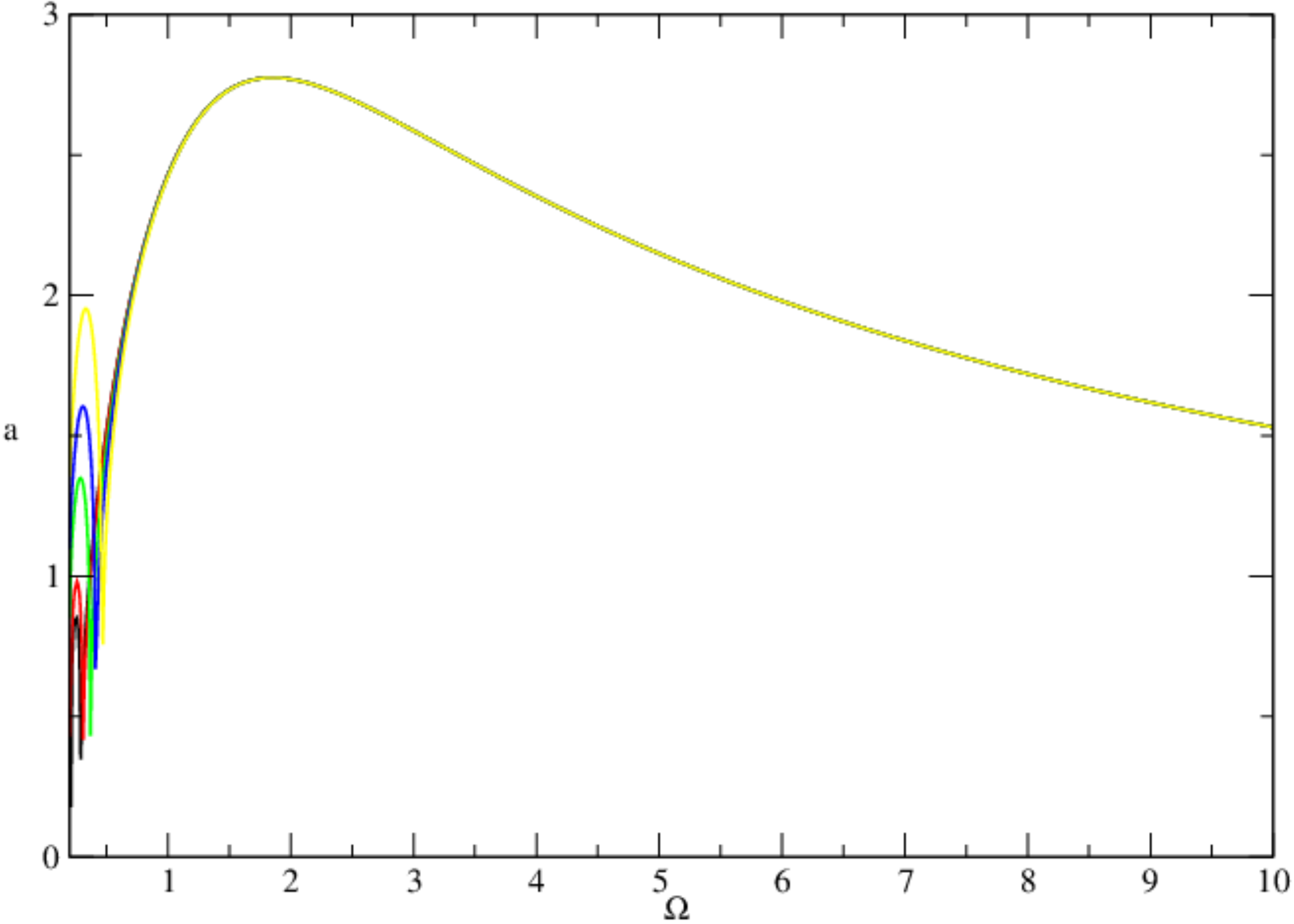}
\end{center}
\caption{Effects of $k_1$ on the frequency-response curves of the order-two subharmonic resonance with the parameters of figure 3.}
\end{figure}

\begin{figure}[htbp]
\begin{center}
 \includegraphics[width=10cm, height=6cm]{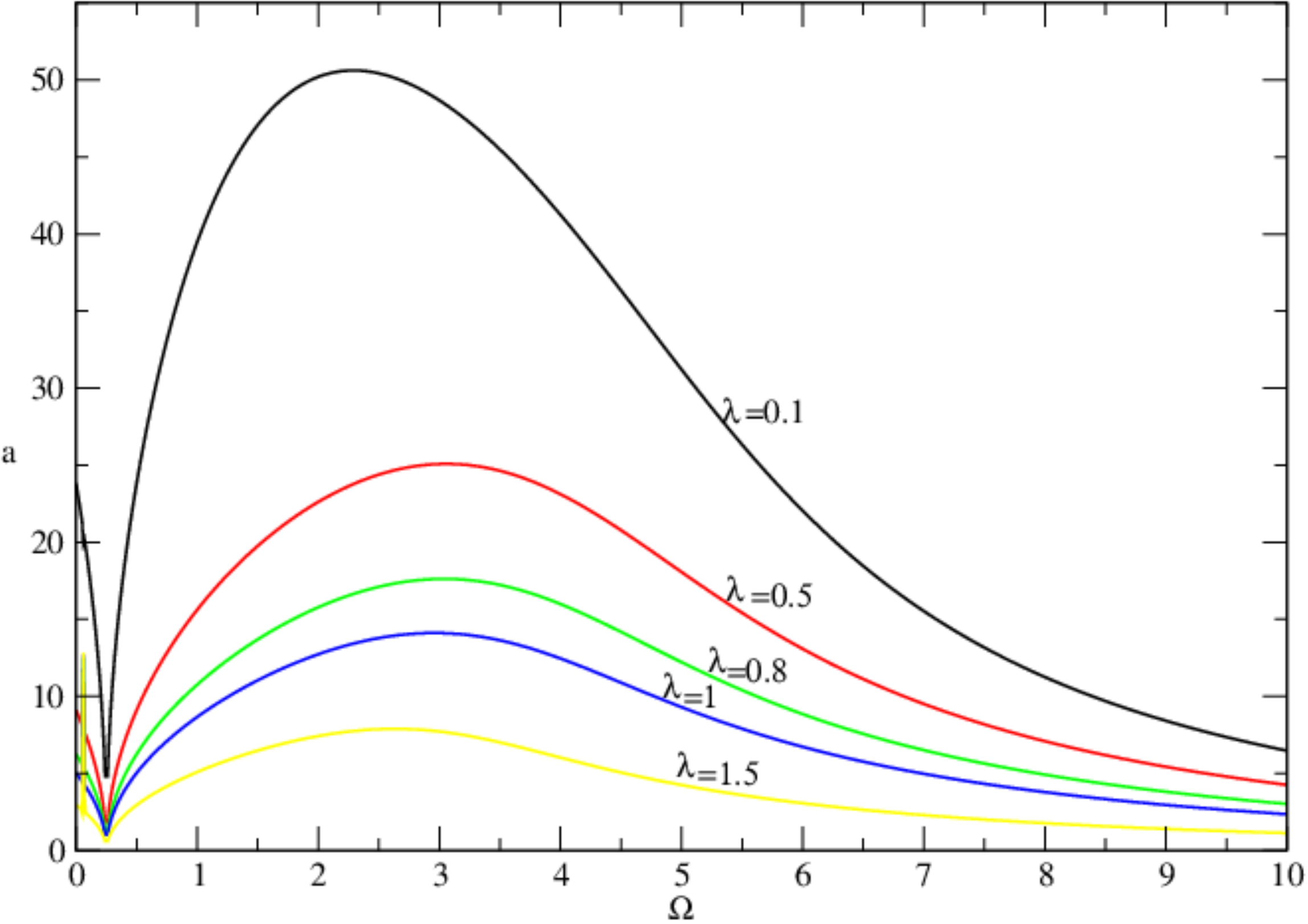}
\end{center}
\caption{Effects of $\lambda$ on the frequency-response curves of the order-three subharmonic resonance with the parameters of figure 4.}
\end{figure}

\begin{figure}[htbp]
\begin{center}
 \includegraphics[width=12cm, height=6cm]{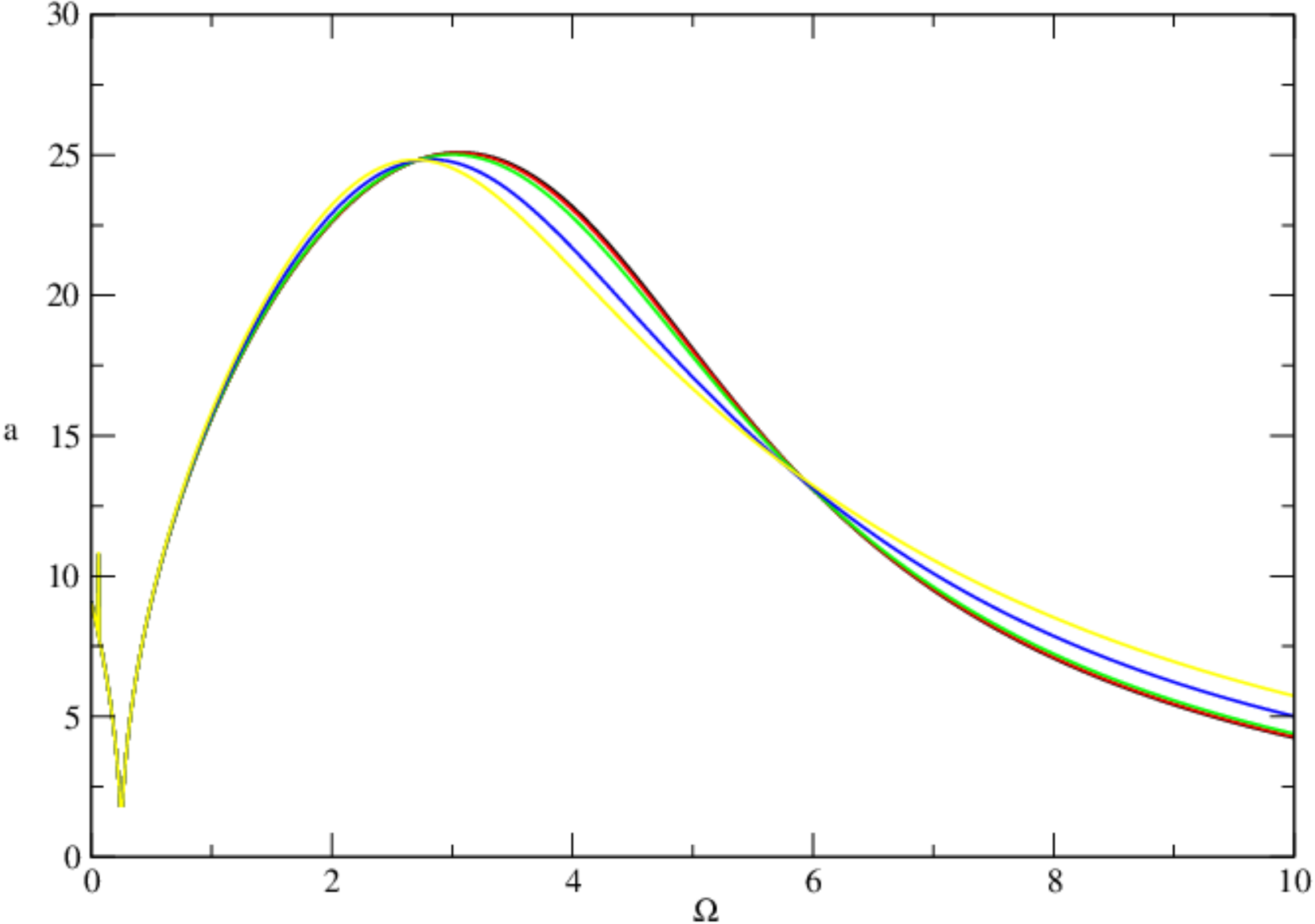}
\end{center}
\caption{Effects of $\mu$ on the frequency-response curves of the order-three subharmonic resonance with the parameters of figure 4.}
\end{figure}

\begin{figure}[htbp]
\begin{center}
 \includegraphics[width=10cm, height=6cm]{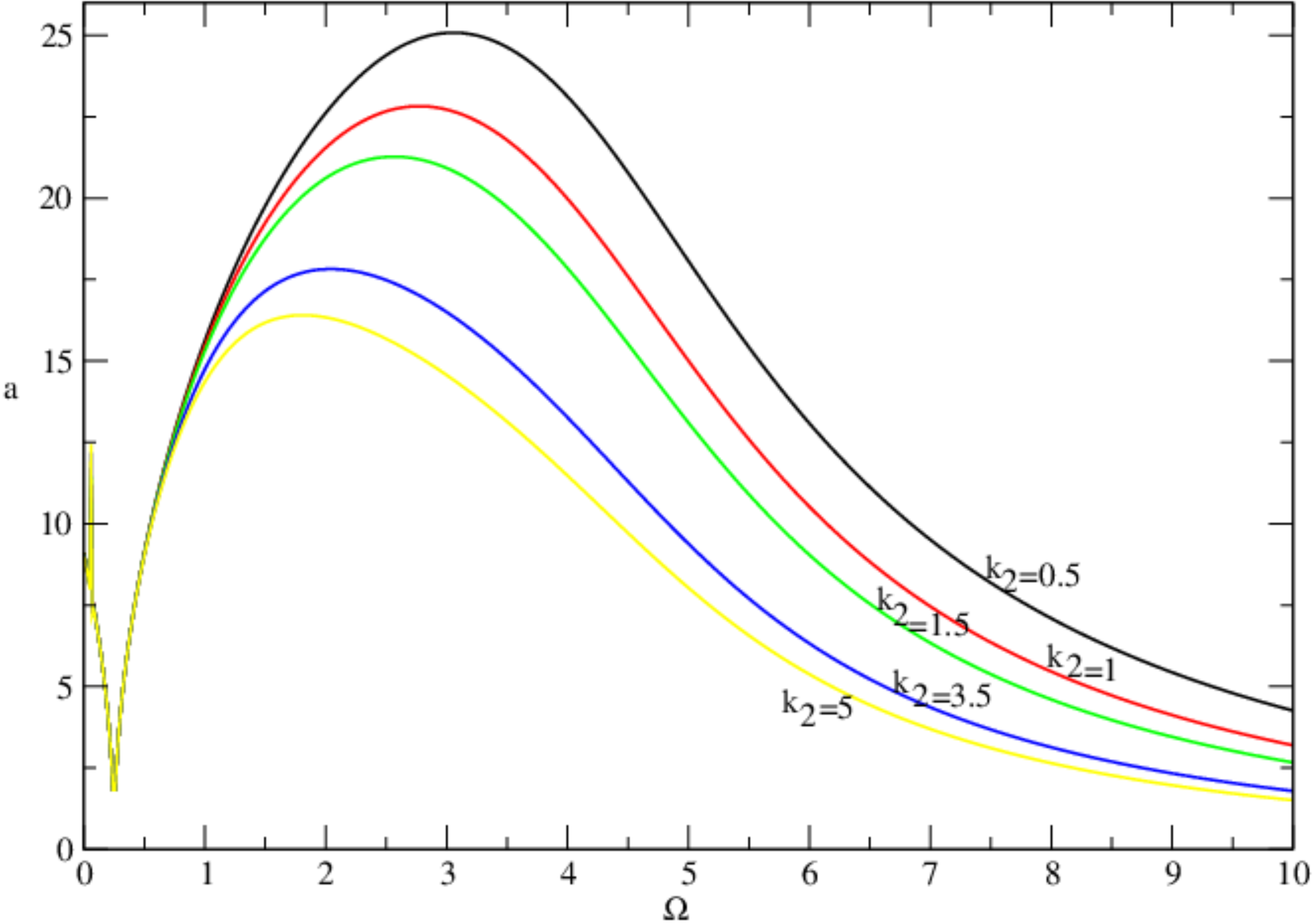}
\end{center}
\caption{Effects of $k_2$ on the frequency-response curves of the order-three subharmonic resonance with the parameters of figure 3.}
\end{figure}
\newpage
\subsection{Effects of parameters on primary resonant state}
In this subsection, we found the effects of parameters $\lambda, \mu, k_2$ and $F$ on primary resonance. Figs. 26-29, show respectively 
the influence of $F, k_2, \lambda$ and $\mu$ on the frequency-response curves of the primary resonance in $(\sigma, a)$ space. From these figure we noticed
that the jump and hysteresis phenomenon are appeared when the amplitude of external excitation and 
or cubic nonlinearities parameters $\lambda$ is increased. We noticed that when  $F$ is increased,
$a$  is increased highly but the variation of $\lambda $ or $k_2$ is not affected the peak value of 
amplitude of vibration in primary resonance. For the dissipation parameter $\mu$, when its is increased
the peak value of $a$ is discreased.     

\begin{figure}[htbp]
\begin{center}
 \includegraphics[width=10cm, height=6cm]{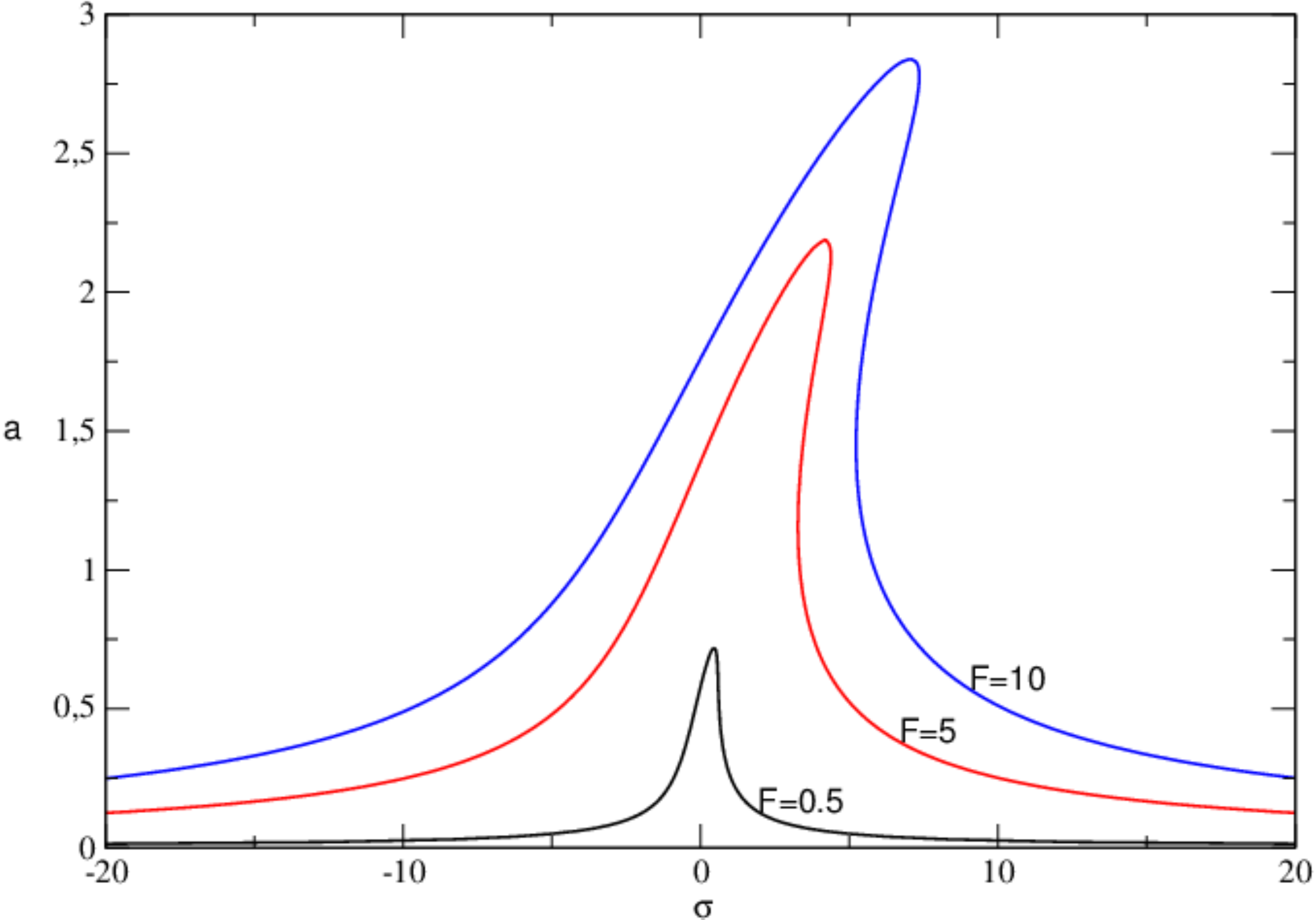}
\end{center}
\caption{Effects of $F$ on the frequency-response curves of primary resonance with $\mu=0.5, \lambda=1, k_2=1$.}
\end{figure}

\begin{figure}[htbp]
\begin{center}
 \includegraphics[width=10cm, height=6cm]{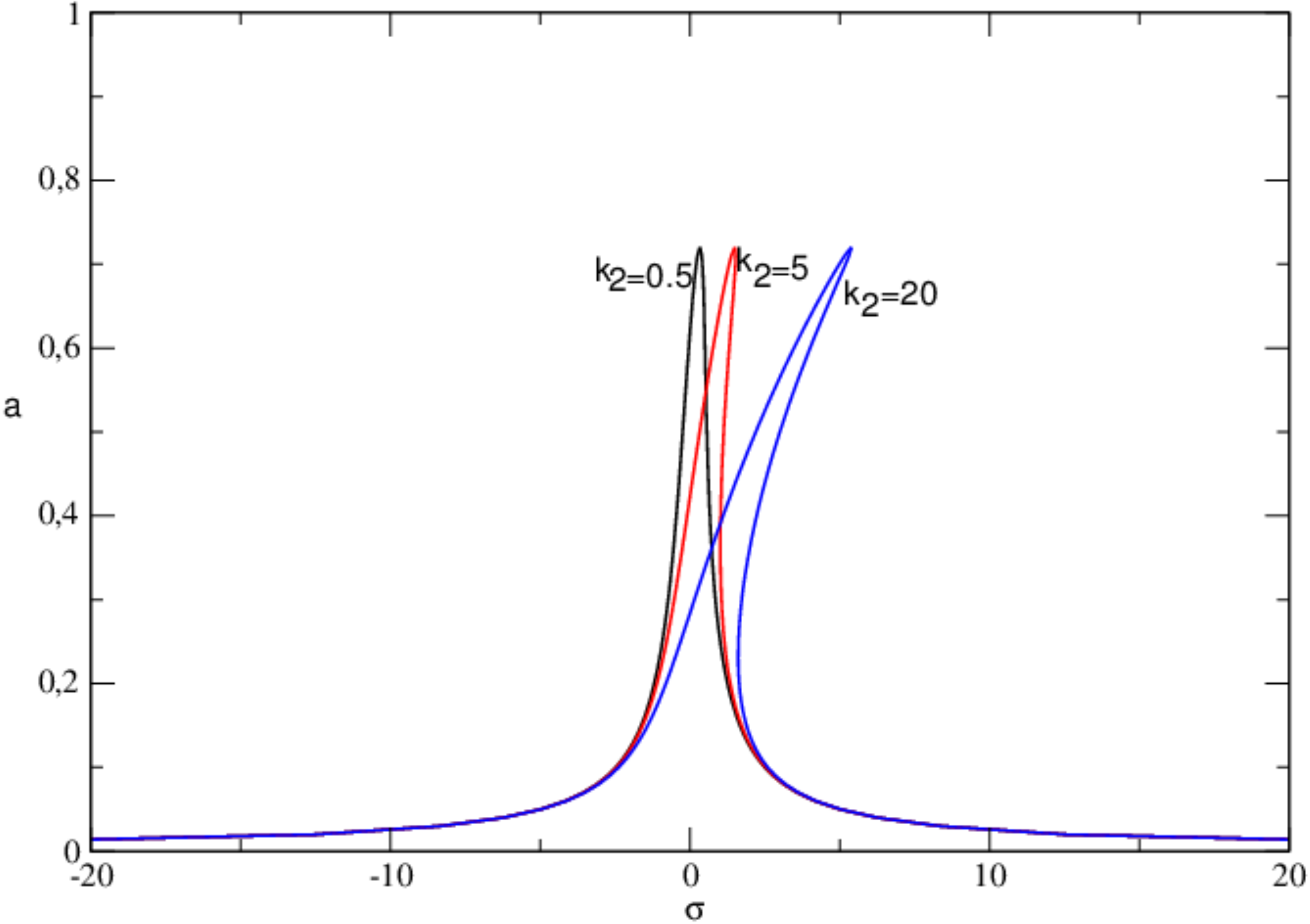}
\end{center}
\caption{Effects of $k_2$ on the frequency-response curves of primary resonance with the parameters of figure 26.}
\end{figure}

\begin{figure}[htbp]
\begin{center}
 \includegraphics[width=10cm, height=6cm]{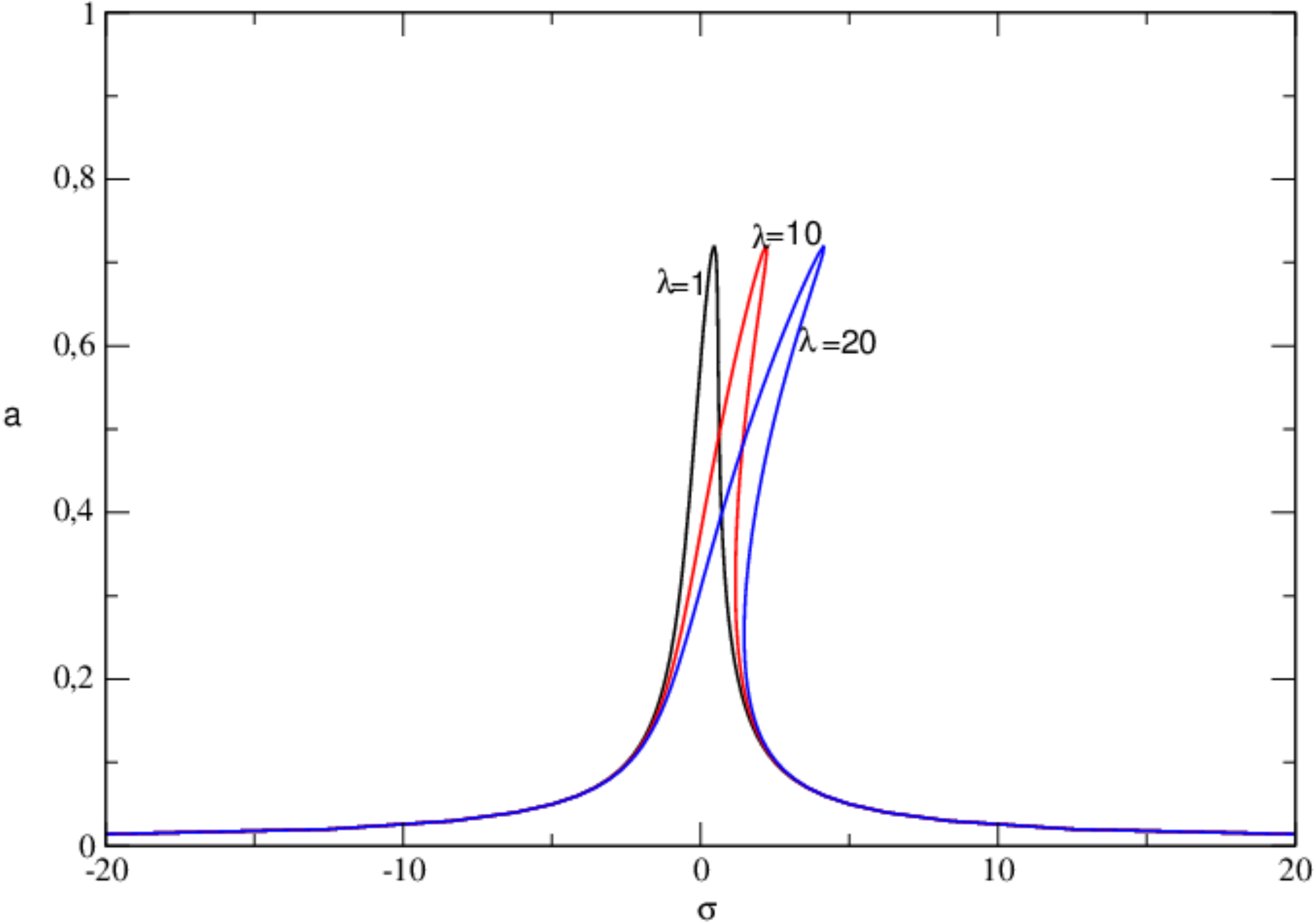}
\end{center}
\caption{ Effects of $k_2$ on the frequency-response curves of primary resonance with the parameters of figure 26.}
\end{figure}

\begin{figure}[htbp]
\begin{center}
 \includegraphics[width=10cm, height=6cm]{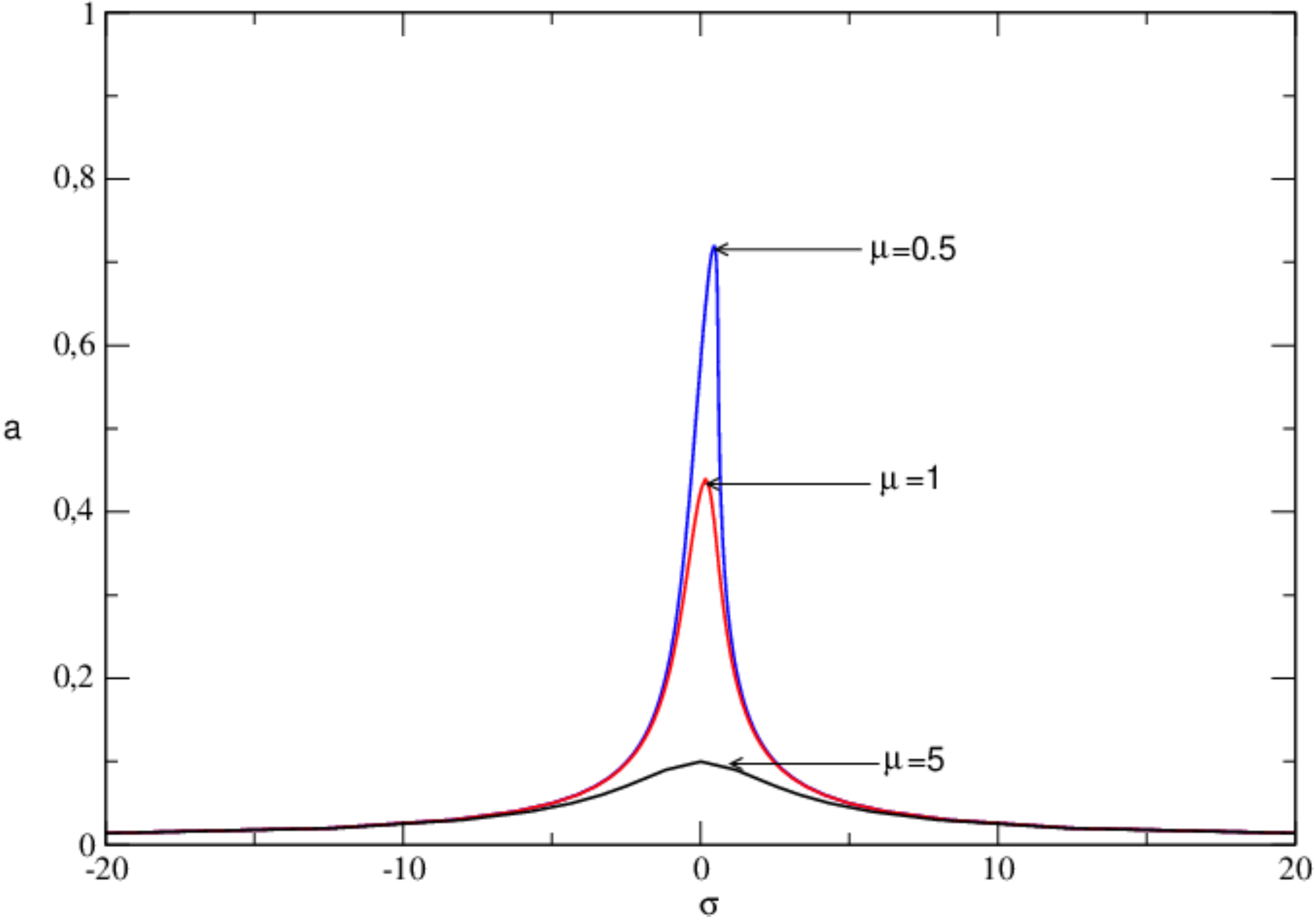}
\end{center}
\caption{Effects of $k_2$ on the frequency-response curves of primary resonance with the parameters of figure 26.}
\end{figure}

\newpage
\section{Discussion}

We have shown the details of a first-order analysis of superharmonic and subharmonic resonances for
the modified Rayleigh-Duffing oscillator. It appears that the curves show antiresonance and resonance peaks for some values
of differents parameters in presence. Through the Figures 1 and 4, it is easy to see that if we take a superharmonic resonance
 curve or subharmonic resonance curve the peak amplitude of resonance increases with the resonance frequency for fixed differents 
parameters of system in the appropriate condition. We also see through the figures(5-10, 11,12) that the cubic nonlinear parameters,
 $\lambda, \mu, k_2 $, and $\sigma$ and $F$
 scales the peak response, while both the quadratic nonlinear parameter and the direct excitation level affect the frequency value of the peak 
response. The superharmonic response of order $\frac{1}{2}$ involves interaction between the parametric excitation and both the nonlinear parameter
 and the direct excitation. At order $\frac{1}{3}$ of this resonance, we also see through the figures 13-17 that the three cubic nonlinear  and $\sigma$ and $F$
parameter have the same effects as the case of order $\frac{1}{2}$. For subharmonic resonance, We also see through the figures(18-25) 
that the cubic nonlinear parameters, $ \mu, k_2 $, scales the peak response, while both the quadratic nonlinear parameter and the direct
 excitation level affect the frequency value of the peak response respectively for two and first appearance of this resonance in order-two or order-three
but $ \lambda$ affect the the peak amplitude and the frequency which his corresponds when the subharmonic resonance appear. The subharmonic
resonance may not be critical to modified Rayleigh-Duffing oscillator. The variation in system responses for changes in some parameters have been 
observed in simulations. In primary resonant state, the vibration amplitude can increasing when cubic nonlinearities parameters are increased. In thise case,
the jump and hysteresis phenomenon can be appeared. These provokes the several variation to the system.  
 In general we note that effects due to cubic nonlinearities on the response curves have a significant
 from physical point of view. 
 It is important to note that around the resonance peaks, the amplitudes and accumulate energies 
of the system device are higher than those received in any oscillations. In this case, this oscillator model can give more
interesting applications in physical or engineering, particulary when the model is used as a MEMS device, Selkov model,
Brusselator etc., but the model with high energies is very dangerous since it can give rise to catastrophe damage. In the 
antiresonance peaks, these systems devices vibrates with small amplitude and accumulates energy. This phenomena is of particular
interst when the model is used as an electromechanical vibration absorber (MEMS
device consisting of a $30 \mu m$ diameter silicon disk which can be made to vibrate by heating it 
with a laser beam resulting in a Hopf bifurcation for example). In other words, for the case of ENSO, resonance peaks correspond
 to a high-temperature water from the ocean. This is very serious because this would facilitate climate change especially regarding
the poor living conditions of certain species of fish. This state of affairs could therefore lead to the disappearance or migration to another place these fish consequences of famine, poverty, fuck the maritime economy etc..
 In the case of peaks of anti-resonance, the phenomenon would be less catastrophic.

\newpage
\section{Chaotic vibration of system}
To illustrate the chaotic vibration of the system dynamics in the resonances regions, simulations
were performed for interesting values of the system parameters, using eq.(\ref{eq.0}) for the forced modified
Rayleigh-Duffing oscillator. The simulations show that the model is highly sentitive to initial conditions,
it can leave a quasi-periodic state for a chaotic state without changing the physical parameters. 
Figs. 30 show the chaotic behavior of the system in subharmonic and superharmonic case but not 
too close to the resonance region. For Figs. 31 and 32 represent the portaits phase of this system 
respectively exactly the  superharmonic and subharmonic resonances. The analysis of Figures 26 shows 
that for given values ​​of the parameters of the system except for the excitation
 frequency, the amplitude of the exciting force and the natural frequency of the system 
which have been selected from the slightly near resonance box, chaotic behavior is not the same.
 The system is less chaotic in the superharmonic case (see Fig. 30 (a), (b) and (c)) than in the
 subharmonic case (see Fig. 30 (d)). These figures also show the influence of the exciting force on 
the intensity of the chaotic behavior of the system. Figures (a) and (b) for the plotted values ​​of 
frequencies and amplitudes of the exciting force at the resonance taken strictly superharmonic order-two 
and order-three, respectively show that the system is quasi-periodic but chaotic in the case of 
figures (a) and (b) obtained under the conditions corresponding to the subharmonic resonance of
order-two and order-three respectively. We note that the quasi-periodicity is less enhanced in the
 case of order-two in the case of order-three for superharmonic resonance and the 
system is less chaotic in the case of order-two in the case of order-three for the subharmonic resonance.
Fig. 33 show the phase portrait of the system at the superharmonic resonance for $\omega=2, \Omega=1 $ (order-two) and $\omega=2, \Omega=1 $ 
(order-three) respectively but the control parameter $\epsilon= 0.0001$. These Fig. 33 ( (a) and (b)) confirm that
the system is quasi-periodic and is less enhanced in the  case of order-two in the case of order-three for this resonance. Indeed by looking at their corresponding time
series, one can observe a quasi-periodic and chaotic states as shown in Fig. 34-35.
Finally, the system is less chaotic in the case of the superharmonic resonance  in the case subharmonic
resonance and in subharmonic and superharmonic case but not too close to the resonance region the system is the very chaotic in thes two cases.
We notice that the chaotic behavior and the intensity of the chaotic behavior of forced modified  Rayleigh-Duffing oscillator depends of the differents parameters 
of the system which influence the differents resonances region. Resonances, quasi-periodicity and chaotic is three phenomenon of this oscillator which depends
between their.

\begin{figure}[htbp]
\begin{center}
 \includegraphics[width=12cm, height=10cm]{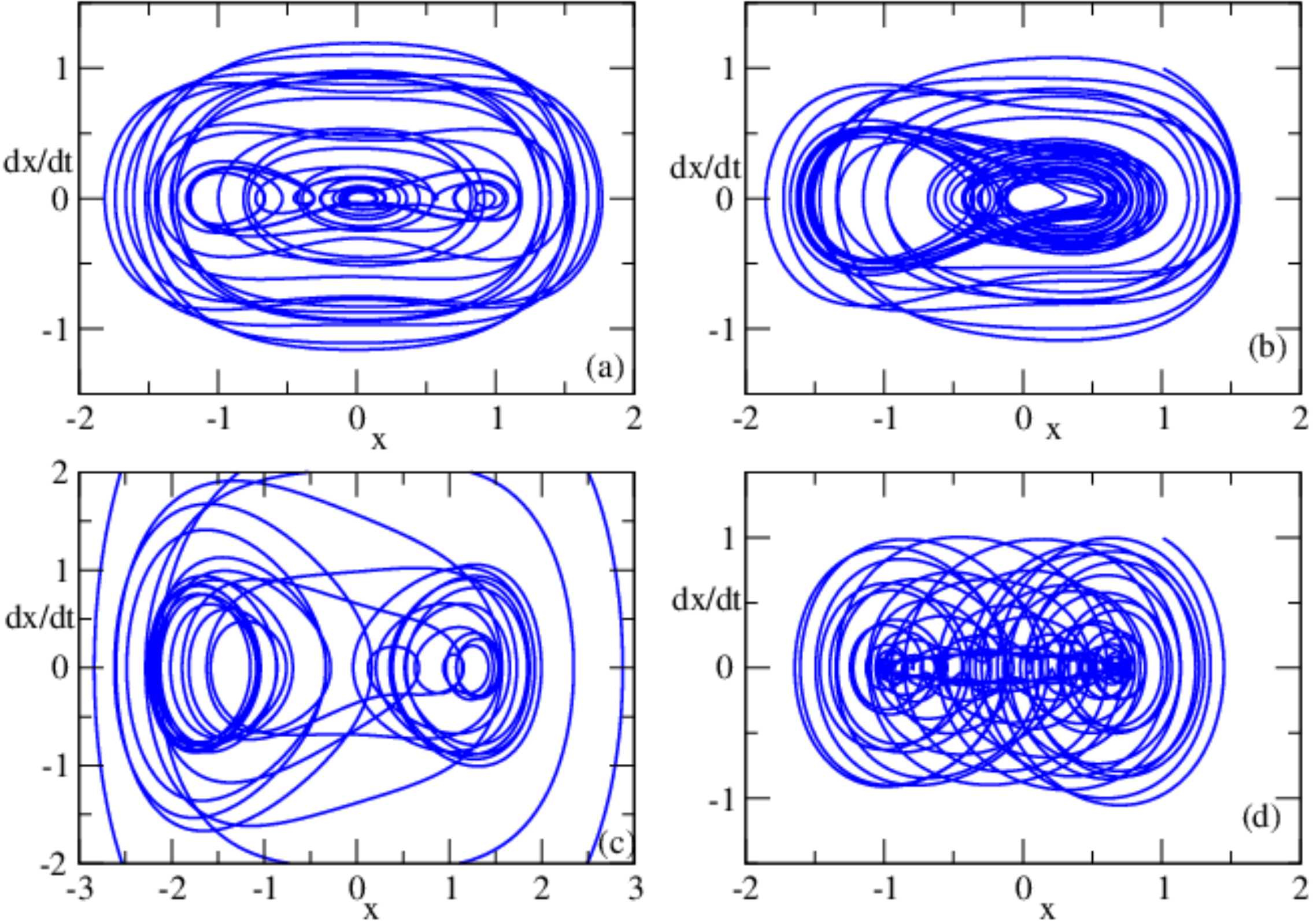}
\end{center}
\caption{Phase portraits of forced modified Rayleigh-Duffing oscillator for $\mu=0.1; \alpha=1; \lambda=1; \beta=0.01; k_2=0.1; k_1=0.1;
 \epsilon=0.1; \Omega=0.1 $.(a) corresponding to $\omega=0.01$ and $F=0.01$, (b) $\omega=0.01$ and $F=0.1$, (c) $\omega=0.01$ and $F=1$, (d) $\omega=1$ and $F=1$.}
\end{figure}

\begin{figure}[htbp]
\begin{center}
 \includegraphics[width=12cm, height=6cm]{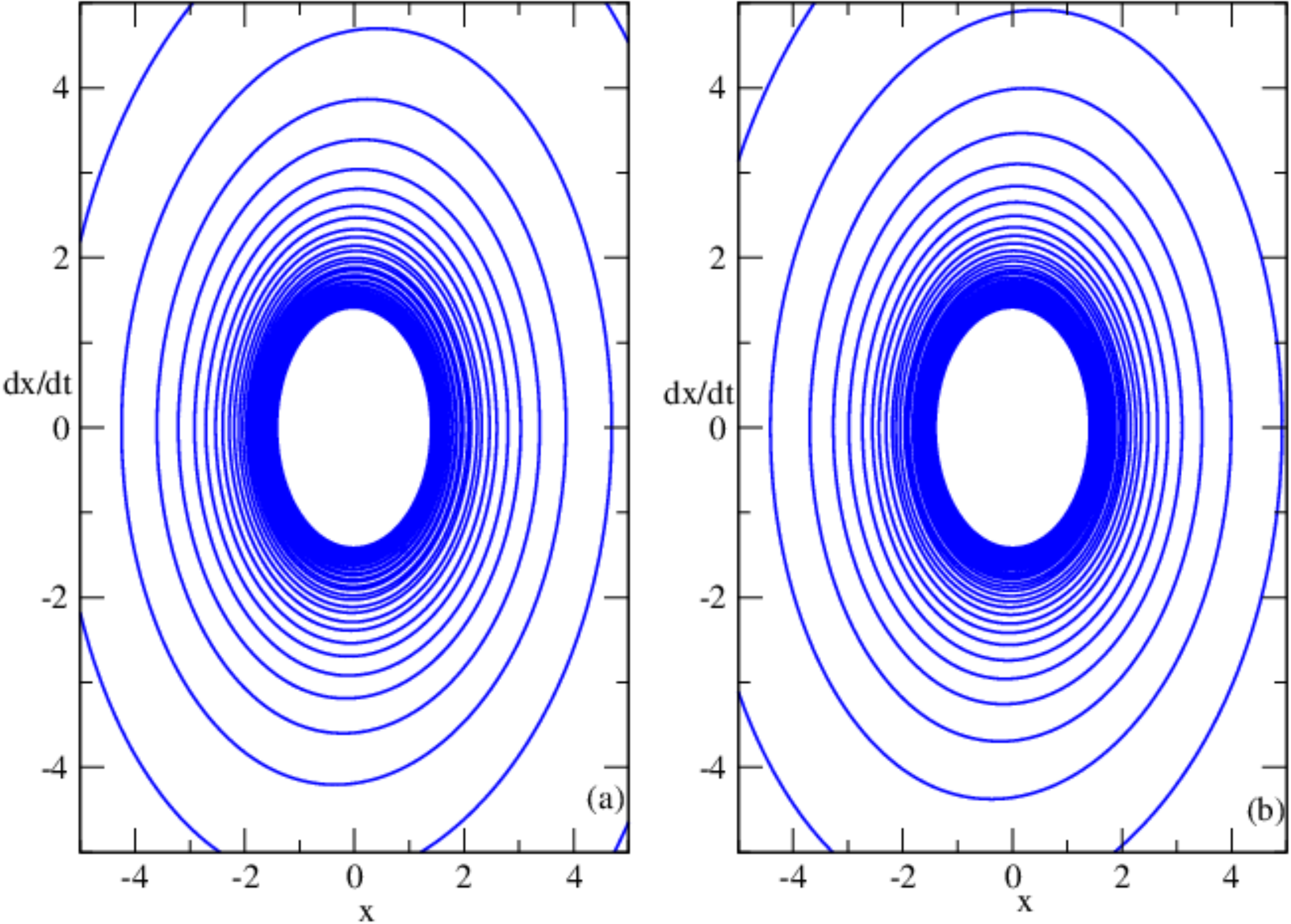}
\end{center}
\caption{Phase portraits of forced modified Rayleigh-Duffing oscillator for $\mu=0.5; \alpha=0.5; \lambda=0.5; \beta=0.1; k_2=0.5; k_1=0.1;
 \epsilon=0.01; F=0.01  $.(a) corresponding to $\omega=1$ and $\Omega=0.5$ and  (b) corresponding to $\omega=1$ and $\Omega=0.3125$}
\end{figure}


\begin{figure}[htbp]
\begin{center}
 \includegraphics[width=12cm, height=6cm]{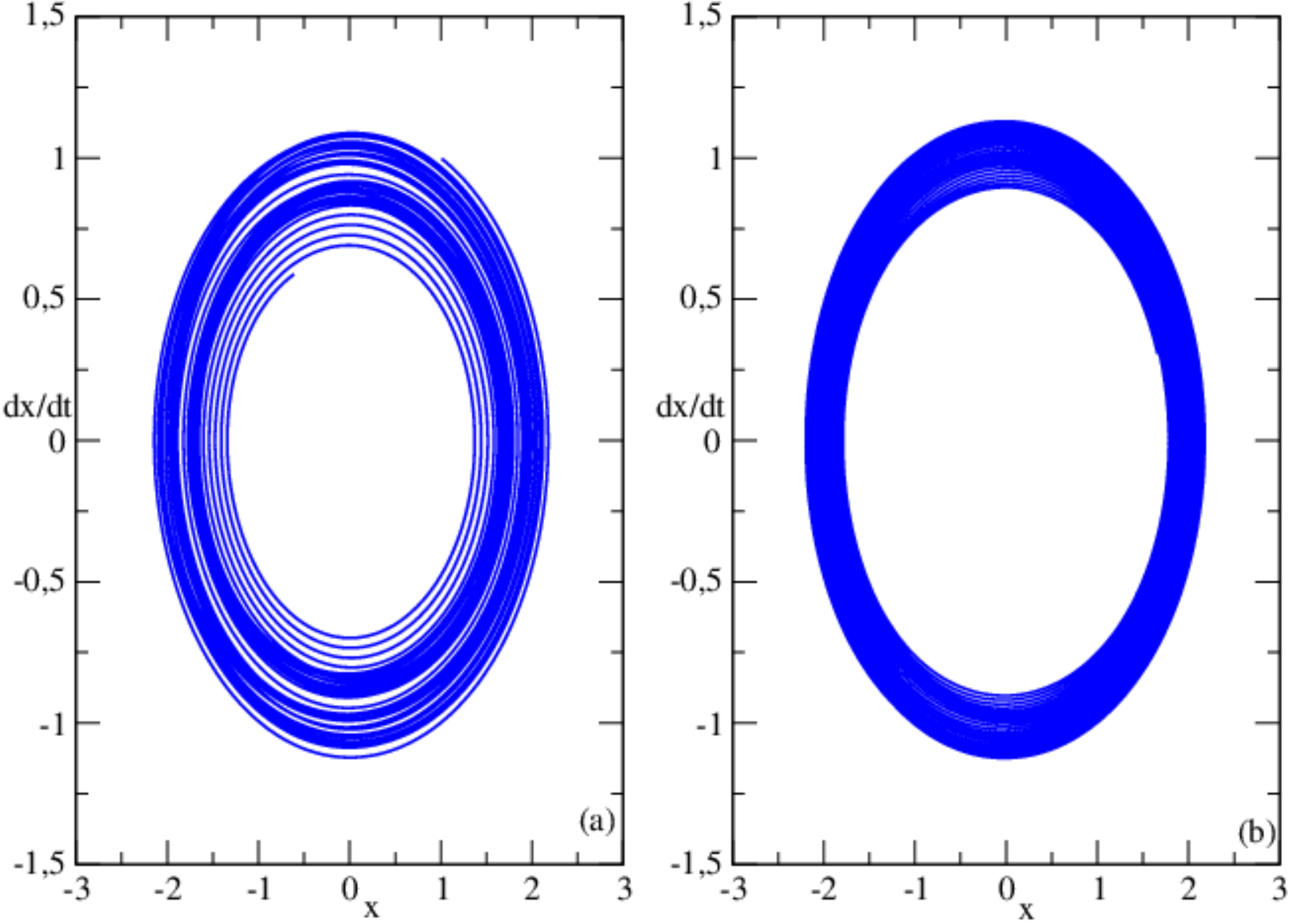}
\end{center}
\caption{Phase portraits of forced modified Rayleigh-Duffing oscillator for $\mu=0.5; \alpha=0.5; \lambda=0.5; \beta=0.1; k_2=0.5; k_1=0.1;
 \epsilon=0.01; F=0.01  $.(a) corresponding to $\omega=0.5$ and $\Omega=1$ and  (b) corresponding to $\omega=0.5$ and $\Omega=1.5$}
\end{figure}

\begin{figure}[htbp]
\begin{center}
 \includegraphics[width=12cm, height=6cm]{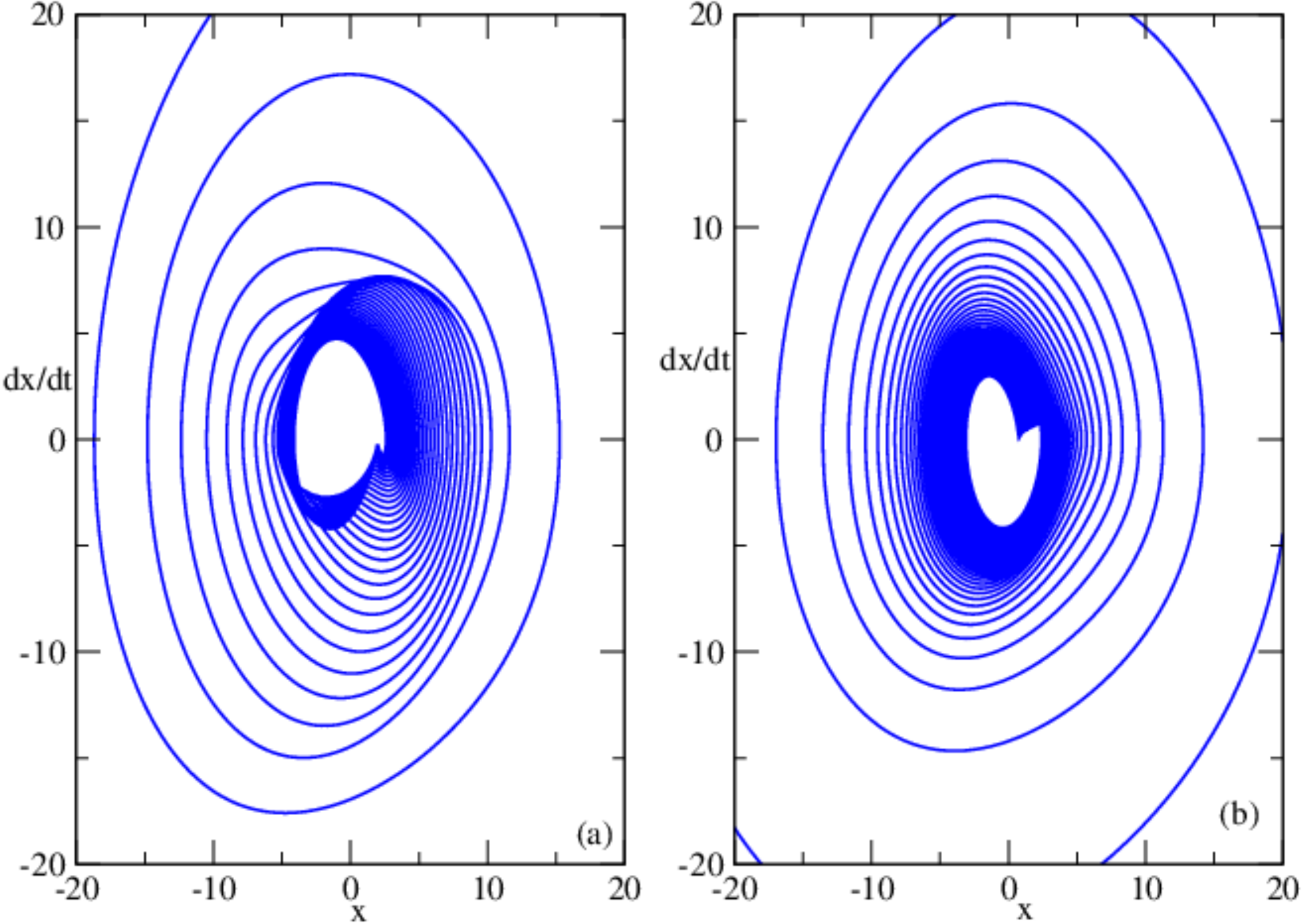}
\end{center}
\caption{Phase portraits of forced modified Rayleigh-Duffing oscillator for $\mu=0.5; \alpha=0.5; \lambda=0.5; \beta=0.1; k_2=0.5; k_1=0.1;
 \epsilon=0.0001; F=0.01  $.(a) corresponding to $\omega=2$ and $\Omega=1$ and  (b) corresponding to $\omega=3$ and $\Omega=1$}
\end{figure}

\begin{figure}[htbp]
\begin{center}
 \includegraphics[width=12cm, height=10cm]{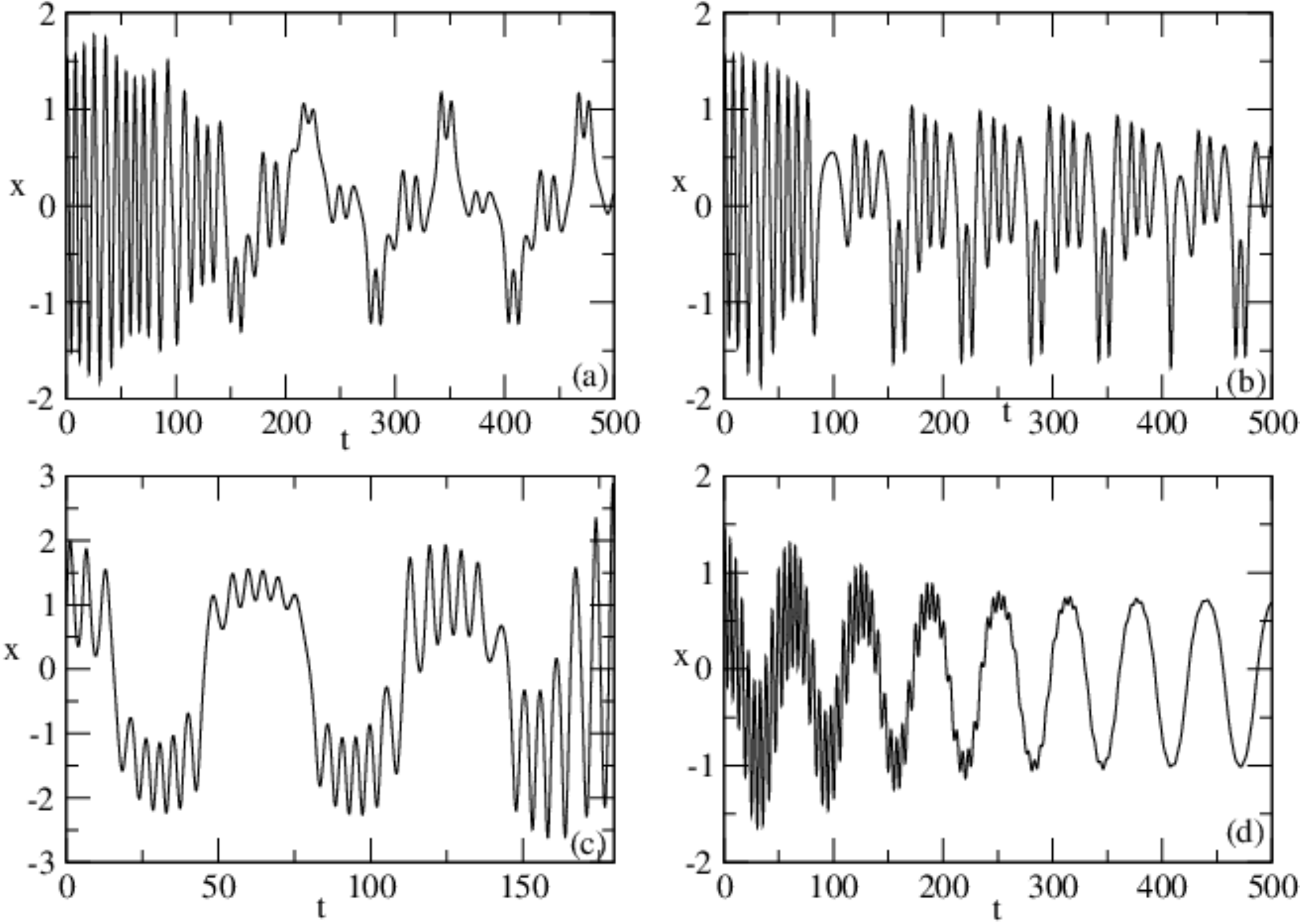}
\end{center}
\caption{Times series showing the quasi-periodic and chaotic states for the superharmonic and subharmonic with the parameters of figure 26.}
\end{figure}

\begin{figure}[htbp]
\begin{center}
 \includegraphics[width=12cm, height=6cm]{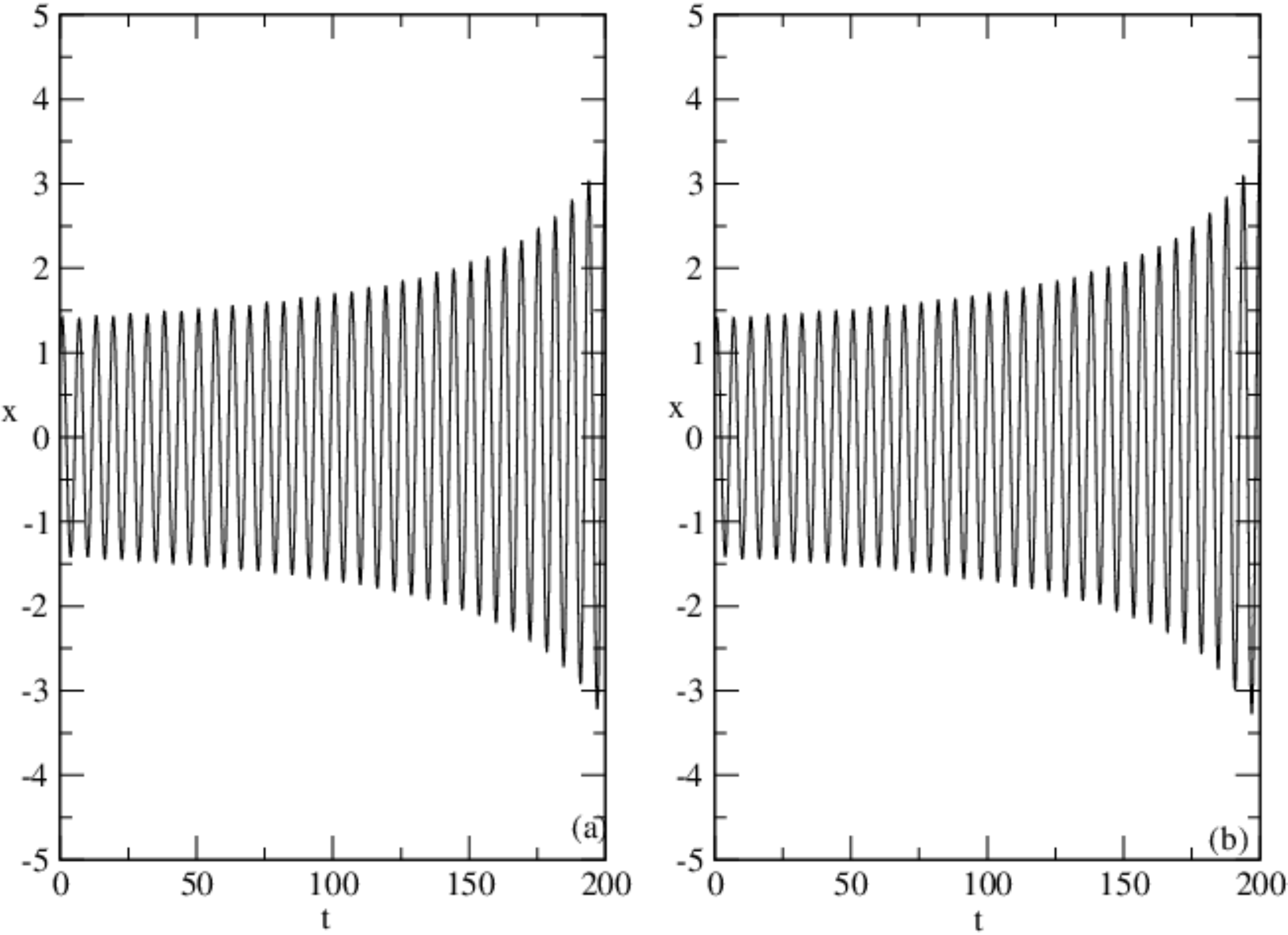}
\end{center}
\caption{Times series showing the chaotic states for the superharmonic resonance with the parameters of figure 27.}
\end{figure}

\begin{figure}[htbp]
\begin{center}
 \includegraphics[width=12cm, height=6cm]{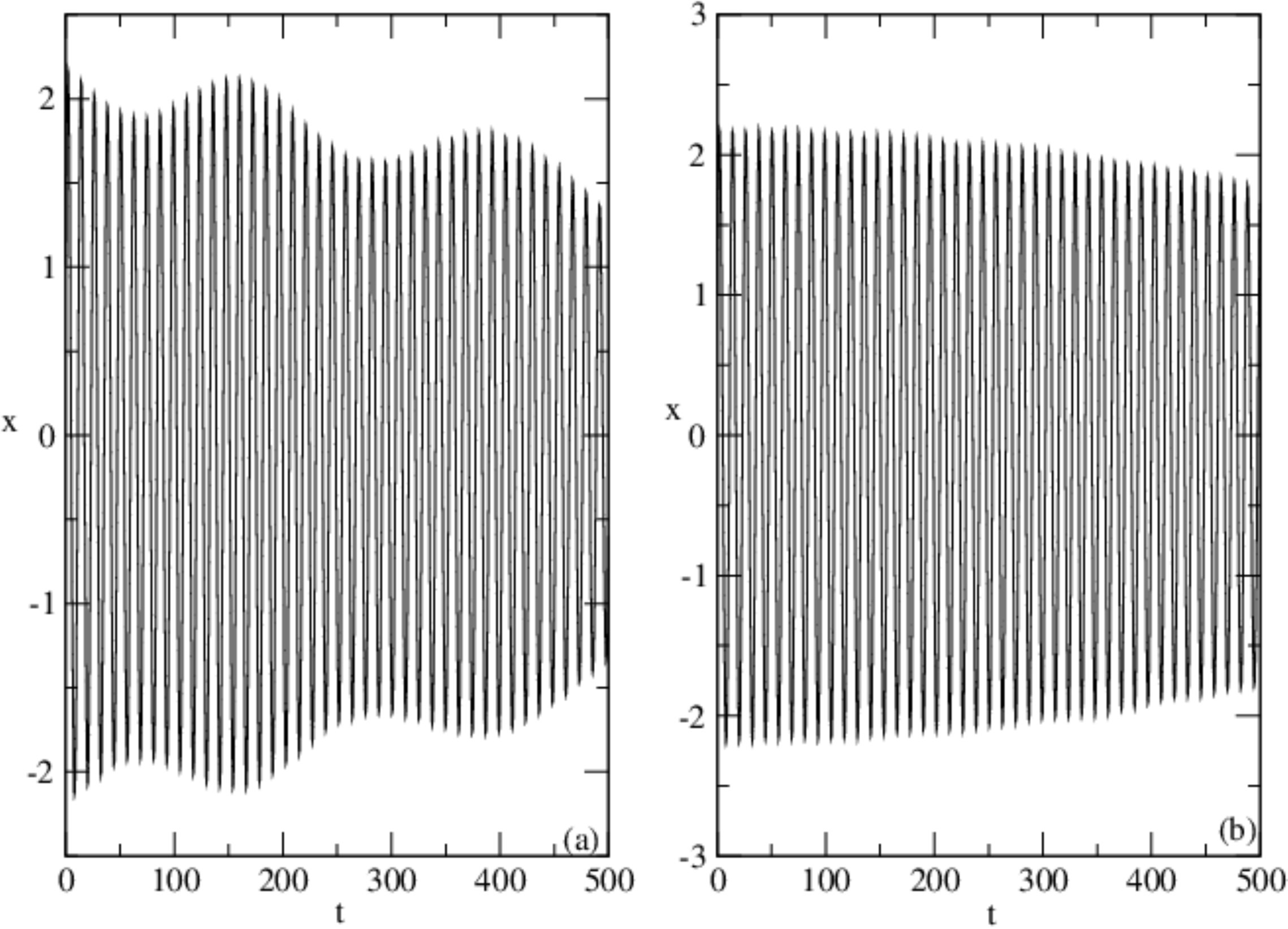}
\end{center}
\caption{Times series showing the chaotic states for the subharmonic resonance with the parameters of figure 28.}
\end{figure}

\begin{figure}[htbp]
\begin{center}
 \includegraphics[width=12cm, height=6cm]{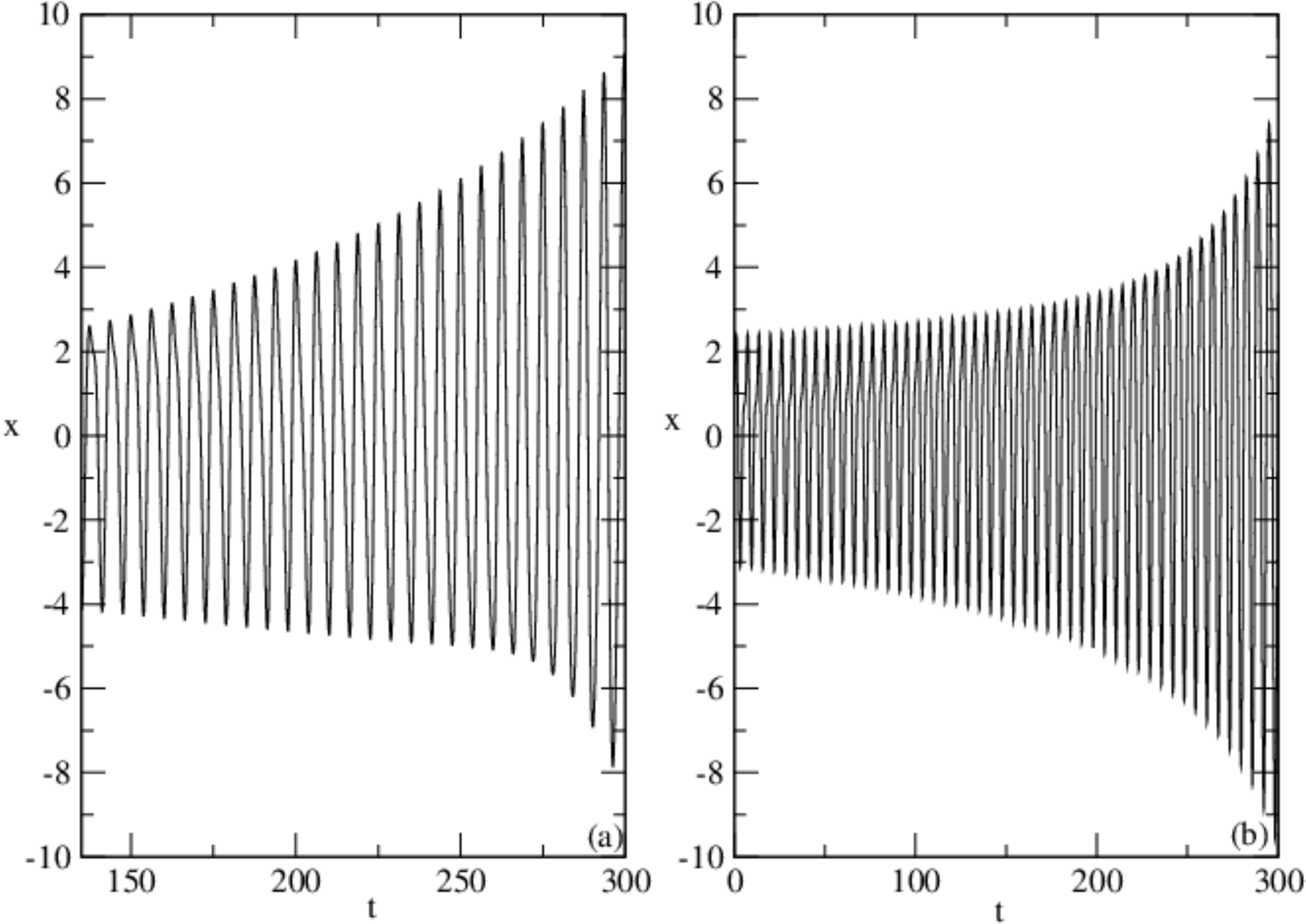}
\end{center}
\caption{Times series showing the chaotic states for the superharmonic resonance with the parameters of figure 29.}
\end{figure}


\newpage
\section{Conclusion}
In this paper, superharmonic, subharmonic and primary resonant sates have been studied. Using the method of multiple scales,
 we obtained the primary resonance and  the order-two and order-three for each type of other resonance. We found also in each case the maximum 
value of the amplitude of the oscillations for the system. We noted that in the case of two-order superharmonic
 or subharmonic resonance, this maximum value depends on all the parameters of the system but in the case of order three,
 only the coefficients of the terms cubic parameters affect the maximum amplitude of the resonance. It should be noted that 
from the simulation of different equations of the resonance curve, the amplitude of the response is higher in the case of all
 subharmonic resonance in the superharmonic. By fixing all the parameters of the system and varying only the amplitude of the 
parametric excitation above the critical value, the increasing amplitude of the parametric excitation provokes a rapid changes
 in the amplitude of the response to the resonances. We obtained the jump and hysteresis phenomenon in the system behaviors.
 The chaotic behavior have study at superharmonic and subharmonic resonances and also 
 subharmonic and superharmonic case but not too close to the resonances regions. We found the differents regions where the systems which modelled by 
the forced modified Rayleigh-Duffing oscillator ( El Ni$\tilde{n}$o Southern Oscillation (ENSO) coupled tropical ocean-atmosphere weather phenomenon, 
 MEMS device  consisting of a $30 \mu m$ diameter silicon disk which can be made to vibrate by heating it with a laser beam resulting in a Hopf bifurcation, 
the modified Selkov equations, modified abstract trimolecular chemical reaction etc.) are quasi-periodic and chaotic.

\section*{Acknowlegments}
The authors thank IMSP-UAC for financial support. C.H. Miwadinou would like to thank Bernard T., Mathias H., Laurent H.
 and Audran K. for their
help in completion of this work.

\end{document}